\documentclass[floatfix,reprint,preprintnumbers,amsmath,amssymb,aps,superscriptaddress,longbibliography,showkeys,nobibnotes]{revtex4-2}

\usepackage[english]{babel}
\usepackage{arydshln}
\usepackage{braket}
\usepackage{graphicx}
\usepackage{placeins}
\usepackage{longtable}
\usepackage{tabularray}
\UseTblrLibrary{booktabs}
\usepackage[colorlinks=true, allcolors=blue]{hyperref}
\usepackage[strings]{underscore}
\usepackage[utf8]{inputenc}
\usepackage{xtab,afterpage}
\usepackage{xcolor}

\newcommand{\beqar}{\begin{eqnarray}}
\newcommand{\eeqar}{\end{eqnarray}}

\begin{document}
\title{Overview of X-ray Thomson scattering measurements of extreme states of matter}

\author{Tobias~Dornheim}
\thanks{Author to whom correspondence should be addressed: \href{mailto:t.dornheim@hzdr.de}{t.dornheim@hzdr.de}}
\affiliation{Institute of Radiation Physics, Helmholtz-Zentrum Dresden-Rossendorf (HZDR), D-01328 Dresden, Germany}
\affiliation{Center for Advanced Systems Understanding (CASUS), Helmholtz-Zentrum Dresden-Rossendorf (HZDR), D-02826 G\"orlitz, Germany}

\author{Hannah~M.~Bellenbaum}
\affiliation{Center for Advanced Systems Understanding (CASUS), Helmholtz-Zentrum Dresden-Rossendorf (HZDR), D-02826 G\"orlitz, Germany}
\affiliation{Institut f\"ur Physik, Universit\"at Rostock, D-18057 Rostock, Germany}

\author{Thomas~Gawne}
\affiliation{Center for Advanced Systems Understanding (CASUS), Helmholtz-Zentrum Dresden-Rossendorf (HZDR), D-02826 G\"orlitz, Germany}

\author{Jan~Vorberger}
\affiliation{Institute of Radiation Physics, Helmholtz-Zentrum Dresden-Rossendorf (HZDR), D-01328 Dresden, Germany}

\author{Dirk~O.~Gericke}
\affiliation{Centre for Fusion, Space and Astrophysics, Department of Physics, University of Warwick, Coventry CV4 7AL, UK}

\date{\today}

\begin{abstract}
Since its first successful applications in the early 2000s, x-ray Thomson scattering (XRTS) has emerged as one of the most successful tools for the diagnostics of extreme states of matter in the laboratory. By sampling the dynamic structure factor of the electrons, XRTS is capable of giving detailed insights into the atomic-scale physics of the matter probed. Moreover, thermodynamic parameters, like the mass density, temperature, and ionization state, are routinely inferred from XRTS measurements, providing a comprehensive characterization of the sample probed. In addition, the dynamic structure factor is of considerable interest in its own right as it contains information on other effects such as the plasmon shift, miscibility between species, electronic states and potential transitions between these states. In this work, we provide an extensive overview of previous XRTS experiments at both traditional laser and X-ray free electron laser facilities, including information about the probed material (elements, conditions), scattering geometry, analysis methods as well as corresponding references. In addition, we briefly discuss the advantages and shortcomings of widely used analysis methods for XRTS spectra and reflect on upcoming future developments in XRTS experiments and theory.
\end{abstract}
\maketitle

\section{Introduction}

Over the last decades, improving our understanding of matter at extremely high densities and temperatures has become a high priority at the interface of condensed matter and plasma physics \cite{fortov_review, wdm_book, vorberger2025roadmapwarmdensematter, bonitz2026quantumeffectsplasmas} with applications in astrophysics, material sciences, and quantum chemistry. 
Arguably, the holy grail for this research is the realization and optimization of inertial confinement fusion \cite{Hurricane_Nature_2014, Betti2016, zylstra_Nature_2022, Hurricane_RevModPhys_2023}.
For this goal, both the fusion fuel and the ablator material have to traverse the highly complex warm dense matter regime \cite{hu_ICF} in a controlled way to reach ignition conditions in the core of highly compressed hydrogen. 
A second source for the extraordinary interest in extreme states of matter is the perpetual quest to understand celestial objects such as giant planet interiors \cite{Militzer_2008,Guillot2018,Brygoo2021,Celliers_Science_2018}, brown dwarfs \cite{becker,saumon1}, white dwarf atmospheres \cite{Kritcher_Nature_2020, SAUMON20221}, and even the crust of neutron stars \cite{Daligault_2009}. The conditions in these objects vary by several orders of magnitude, but they are mostly in a complex quantum state with strong interactions between the particles.
At the lower end of excitation strength, material science and discovery is a highly active frontier \cite{Kraus2016,Kraus2017,Ramakrishna_JPhysCondMat_2020,Lazicki2021}. The focus is here on controlled modifications of material properties through the creation of transient high-pressure states.

Driven by these applications, extreme states of matter are nowadays routinely created in many research facilities all around the globe using a variety techniques to heat and compress different materials, see, e.g., the overview by Falk~\cite{falk_wdm} and the extensive roadmap by Vorberger \textit{et al.}~\cite{vorberger2025roadmapwarmdensematter}.
A key challenge these experiments is the accurate diagnostics of the states created \cite{Riley_book, Pascarelli_NatureReviews_2023, LANDEN2024101102, Kraus_NatureReviewPhysics_2026}. Applied techniques must be suited for high pressures, densities, and temperatures as well as the ultra-fast timescales these highly excited systems evolve(usually $10^{-12}$ to $10^{-9}\,$s).

For over a quarter of a century, XRTS has been established as a powerful in-situ tool for the diagnostics of extreme states of matter \cite{LANDEN2024101102}.
This diagnostic approach is, in principle although not necessarily in practice, capable of dealing with all the challenges these states present \cite{siegfried_review}.
In short, the probability of the scattering for the incoming radiation is directly related to the electronic dynamic structure factor $S_{ee}(q,\omega)$, where $\hbar q$ and $\hbar\omega$ correspond to the momentum transfer and energy loss of the scattered photon, respectively~\cite{CROWLEY201455,sheffield2010plasma,acosta2026montecarloeventgenerationxray}. The dynamic structure factor can be expressed in various equivalent ways \cite{sheffield2010plasma,quantum_theory}, e.g.,
\begin{eqnarray}
    S_{ee}(q,\omega) &=& \frac{1}{2\pi} \int\limits_{-\infty}^\infty 
                                        \textnormal{d}\omega\ F_{ee}(q,t)\ e^{i\omega t} 
    \label{eq:F}
    \\
    &=& \sum_{l,m} P_m \left|\bra{l} \hat{n}_e(\mathbf{q})\ket{m}\right|^2 
                                    \delta\!\left(\omega - \frac{E_l-E_m}{\hbar}\right) \,. 
    \nonumber\\
    \quad
    \label{eq:spectral_representation}
\end{eqnarray}
Eq.~(\ref{eq:F}) defines the dynamic structure factor as the Fourier transform of the intermediate scattering function $F_{ee}(q,t)=\braket{\hat{n}_e(q,t)\hat{n}_e(-q,0)}$, which measures temporal correlations within the single electron density $\hat{n}_e(q,t)$ in momentum or wavenumber space.
Additional insight into the physics encoded by the dynamic structure factor can be obtained from Eq.~(\ref{eq:spectral_representation}). This spectral representation is given by the sum over all possible transitions between electronic eigenstates $m$ and $l$ with associated eigenenergies $E_m$ and $E_l$ and occupation probabilities $P_m$.
From Eqs.~(\ref{eq:F}) and (\ref{eq:spectral_representation}), it becomes immediately clear that XRTS gives fairly detailed information about the electronic structure and a wealth of other electronic and ionic properties of the probed sample. This includes both the extraction of basic plasma parameters such as density, temperature, and ionization degree as well as detailed physics effects such as collisional damping of collective electron excitations \cite{Sperling_PRL_2015}, the miscibility of different elements \cite{Frydrych_NatComm_2020} or ionization potential depression \cite{Bellenbaum_PRR_2025,kraus_xrts}.

The state-of-the-art of XRTS experiments has come a long way since the pioneering efforts by Landen, Gregori, Glenzer, Riley and others \cite{LANDEN2001465,Glenzer_PRL_2003,Gregori_AIP_2002,Gregori_PRE_2003,Gregori_POP_2004,Glenzer_PRL_2007,Riley_IEEE_2003,Riley_2007,Neumayer_RSI_2006} in the early and mid-2000s.
For example, XRTS measurements are nowadays routinely deployed for diagnosing imploding capsules by x-rays from laser-driven plasmas at the OMEGA laser facility in Rochester \cite{Fletcher_POP_2013, Fletcher_PRL_2014, Chapman_NatComm_2015, KRITCHER2011271, Kritcher_PRL_2011} and the National Ignition Facility (NIF) in Livermore \cite{Chapman_POP_2014, Kraus_PRE_2016, Tilo_Nature_2023, Dornheim_NatComm_2025},
resulting for example in the observation of the onset of K-shell ionization in strongly compressed beryllium by D\"oppner \textit{et al.}~\cite{Tilo_Nature_2023}.

X-ray free electron laser (XFEL) facilities --- starting with facilities for VUV and soft x-ray light such as FLASH in Hamburg \cite{HOLL2007120,Toleikis_2010} and continuing with sources for hard x-rays such as the Linac Coherent Light Source (LCLS) and the European XFEL --- constitute another game changer for the field. These XFELs provide light that is much brighter than any laser-driven source and provide very short pulses of photons with a narrow spectral bandwidths directed into a very small solid angle. This strongly reduces background noise from the source and also defines the scattering geometry much better compared with laser-driven plasma sources. XFELS can also be combined with high repetition optical lasers which allows pump-probe experiments \cite{Zastrau_JSynchRad_2021, Gorman_POP_2024, bespalov2026experimentalevidencebreakdownuniformelectrongas} with unprecedented photon statistics and, thus, very high data quality. It has been shown that the spectral properties of XFELs can be further improved by seeding the x-ray emission (see, e.g., Fletcher \textit{et al.} \cite{Fletcher2015}). 

The most recent milestone we want to mention here is the development of a 4-pass silicon monochromator that further reduces the spectral width of the probe radiation and diced crystal analyzers (DCA) for catching the scattered photons with high efficiency \cite{McBride_RSI_2018, Wollenweber_RSI_2021}. Combining both techniques allows for XRTS measurements with meV resolution, which opened the possibility for probing ionic modes \cite{White_PRR_2024,Descamps_SciReports_2020}, a direct determination of the ion temperature \cite{White_Nature_2025}, and investigations of plasmon excitations on electronic energy transfer scales \cite{Gawne_PRB_2024,Gawne_ElectronicStructure_2025,gawne2025orientationaleffectslowpair}.

In light of the exciting developments listed above and the ever increasing diversity of applications of XRTS in a plethora of contexts, we feel that a dedicated overview of previous experimemts employing XRTS is overdue.
As such, the present work is not meant as a complete review article, especially since specific aspects of XRTS have been thoroughly discussed already \cite{siegfried_review,Dornheim_review,vorberger2025roadmapwarmdensematter,Kraus_NatureReviewPhysics_2026}. Instead, we strife to give a broad overview about XRTS experiments at different kinds of facilities 
in a great variety of setups, tabulating the most important parameters such as the probed material, scattering angle and beam energy, the extracted conditions and the analysis methods applied. 
In the discussion, we will focus on important milestones and advances that provide the basis for further developments and, thus, will remain relevant for future campaigns. It is our hope that this collection will prove to be useful for newcomers and seasoned practitioners of XRTS alike.

Our overview is organized as follows. In Sec.~\ref{sec:theory}, we briefly summarize some theoretical aspects of XRTS and provide references to the most important techniques and simulation methods.
Sec.~\ref{sec:overview} contains the heart of the present work: a collection of over 90 XRTS experiments at 11 different facilities probing a plethora of different materials at a very broad range of conditions.
For some cases, we also provide links to further analysis giving an alternative interpretation of the experimental data.
The paper is concluded by a summary and outlook in Sec.~\ref{sec:outlook}.

\begin{figure*}\centering
\includegraphics[width=0.7495\textwidth]{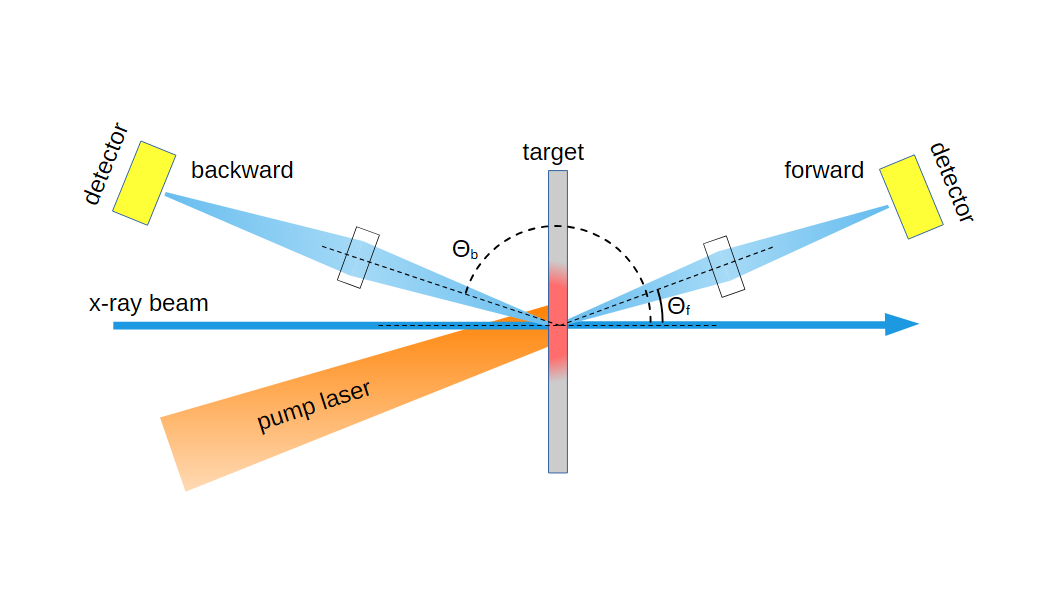}
\caption{\label{fig:fig:fig} 
Schematic overview of a typical x-ray Thomson scattering experiment to diagnose extreme states of matter. The pump laser heats (and compresses) the target. The x-ray probe beam is focused on the center of the heated region of the target. Several x-ray detectors are placed at different scattering angles to obtain the energy resolved scattering spectra. Typical geometries featuring small scattering angles in forward direction allow detection of collective effects. Larger scattering angles (in particular in backward direction) collect information on single particle properties.
}
\end{figure*}

\section{XRTS theory}\label{sec:theory}

\subsection{Connecting XRTS theory and experiments}
\label{sec:intensity}

The measured intensity in XRTS experiments (see Fig.~\ref{fig:fig:fig} for a schematic set-up) is not only determined by the scattering probability but also by details of the x-ray generation and their transport to the detector, e.g., the x-ray source spectrum, multiple processes of x-ray absorption on the way to, in and out the target probed, as well as the detector efficiency.
From a theoretical perspective, it is often expressed as a convolution of the dynamic structure factor $S_{ee}(q,\omega)$ \eqref{eq:F}, 
which contains the physics information that we actually care about, with the combined source and instrument function (SIF) $R(\omega)$
\begin{eqnarray}\label{eq:intensity}
    I(q,\omega) = S_{ee}(q,\omega_0-\omega) \circledast R(\omega) \ ,
\end{eqnarray}
where $\hbar\omega_0$ and $\hbar\omega$ denote the central x-ray probe energy and energy shift, respectively.
The SIF can be expressed as
\begin{eqnarray}\label{eq:SIF}
    R(\omega) = \Sigma(\omega) \circledast \Gamma(\omega)\ ,
\end{eqnarray}
with $\Sigma(\omega)$ and $\Gamma(\omega)$ being the x-ray source function and detector function, respectively.
In general, Eq.~(\ref{eq:intensity}) constitutes an approximation for at least two reasons. First, the momentum transfer $\hbar q$, which is computed from the scattering angle $\theta$, can only be treated as constant when $\omega_0\gg\omega$. This is generally fulfilled for hard x-ray sources and/or relatively narrow spectral windows, e.g., in forward scattering geometry \cite{siegfried_review, Dornheim_T2_2022}. 
Otherwise, the LHS of Eq.~(\ref{eq:SIF}) would need to be re-written as $I(q(\omega),\omega)$, which would make the analysis considerably more complicated and which would also destroy useful symmetry properties such as the detailed balance relation \cite{quantum_theory}.
Second, even the treatment of Eq.~(\ref{eq:intensity}) as a comparably simple convolution constitutes an approximation as, in principle, the detector function $\Gamma(\omega)$ and, hence, the SIF $R(\omega)$ depend on two energy arguments, e.g., the transfer frequency $\omega$ and the incoming probe frequency $\omega_0$. In other words, the SIF depends on the particular energy bin of the XRTS detector, transforming Eq.~(\ref{eq:intensity}) into a significantly more complicated kernel integral equation; see the recent work by Gawne \textit{et al.}~\cite{Gawne_JAP_2024} for an accessible discussion.
Ultimately, the ideal way to describe the measured intensity is given by a dedicated ray-tracing simulation~\cite{Gawne_CompPhysComm_2026}, which automatically takes into account all of the upper effects.
Moreover, ray-tracing can be used to generate an accurate SIF given the x-ray source profile $\Sigma(\omega)$ and the specifications of the detector.

Let us conclude this perspective on how to connect theory and the measured XRTS intensity with a concise list of other subtleties:

In many experimental setups, the target might not be homogeneous. In that case, the detected intensity Eq.~(\ref{eq:intensity}) will be given by a weighted sum over contributions from different conditions \cite{Fortmann_HEDP_2009, Thiele_PRE_2010, Sperling_2013, Chapman_POP_2014, Golovkin_HEDP_2013, Poole_PPCF_2025}. Very steep gradients in the system might even directly affect the form of the Thomson scattering cross section \cite{Kozlowski_SciRep_2016, Belyi_SciRep_2018, Bornath_2019}.

The finite size of the probed target can also have a number of effects by itself, including a superposition of photons on the detector that have been scattered under a range of different scattering angles $\theta$ as well as x-ray attenuation effects~\cite{Chapman_POP_2014}.

Similarly, the finite size of the photon detector results in a corresponding finite angular coverage, which, in turn, leads to an effective superposition of different scattering contributions of a given $q$-window, see, e.g., Refs.~\cite{Gawne_PRB_2024, Gawne_ElectronicStructure_2025, gawne2025orientationaleffectslowpair}.

The target should also be small enough that an x-ray is not scattered multiple times on its path through the target. Due to the very small Thomson cross section, this constraint is usually fulfilled. 

A central assumption of interpreting XRTS spectra is a sufficiently weak photon flux, such that the effect of x-ray absorption on the target can be neglected. This is generally a safe assumption for pump-probe experiments, but does not hold for isochoric heating experiments where the XFEL is used as both pump and probe at the same time \cite{Faustlin_PRL_2010, Sperling_PRL_2015,kraus_xrts, Bellenbaum_APL_2025}. The practical impact of such nonequilibrium effects can be severe and is still not fully understood \cite{Chapman_PRL_2011, Bellenbaum_APL_2025}.
 
Some XRTS measurements, e.g., on optically pumped targets, might be afflicted with an additional radiation background that overlays the scattering spectrum \cite{Fortmann_HEDP_2006}. Moreover, strongly radiative plasmas can also change the scattering probabilities directly \cite{Cross_2016}.

To summarize, the theoretical interpretation of XRTS measurements is generally fairly complex and, thus, should consider many experimental details. Nevertheless, the simplified form of Eq.~(\ref{eq:intensity}) often holds to a very high degree, making it sufficient for practical analysis and diagnostics.

\subsection{Models for the dynamic structure factor}
\label{sec:DSF}

In most cases, the convolution (\ref{eq:intensity}) cannot be straightforwardly inverted,
which, of course, also applies for the more involved full kernel extension \cite{Gawne_JAP_2024}.
This limitation arises mostly due to the combination of experimental noise, finite photon count, and the limited detector range.
Therefore, the interpretation of XRTS measurements has been performed mostly on the basis of \emph{forward modeling}, where theoretical results for $S(q,\omega)$ are convolved with a SIF and then compared to the experimental data.
In particular, a-priori unknown plasma parameters such as the mass density, temperature and ionization state are treated as free fit parameters and can then be inferred from a best fit to the experimental intensity signal, see Refs.~\cite{LANDEN2001465, Gregori_PRE_2003} for the original idea.
A more complete approach is given by additional Markov Chain Monte Carlo sampling, which also provides a reasonable estimate for the uncertainty of the extracted parameters \cite{Kasim_POP_2019}.

Over the decades, a multitude of theoretical models for the dynamic structure factor have been developed with different levels of sophistication \cite{Dornheim_review, siegfried_review, vorberger2025roadmapwarmdensematter}.
In the following, we briefly list the main branches of the theoretical description and corresponding references.

Gregori \textit{et al.}~\cite{Gregori_AIP_2002, Gregori_PRE_2003} have suggested to work within the chemical picture that labels electrons as either bound or free. Then one can apply the Chihara decomposition for the dynamic structure factor \cite{Chihara_1987}, which yields three distinct contributions: (a) transitions between two free electron states (free-free); (b) transitions, where a bound electron is lifted into the continuum (bound-free); and (c) the quasi-elastic feature due to scattering from tightly bound electrons and the screening cloud that both follow the much slower ionic motion. Recently, B\"ohme \textit{et al.} \cite{bohme2023evidencefreeboundtransitionswarm} have convincingly argued that, at a finite temperatures and in thermal equilibrium, the exact detailed balance relation mandates the presence of (d) free-bound transitions, where an initially free electron is de-excited into an energetically lower bound state by the scattering photon. The distinction between bound and free electrons within the Chihara approach is not always straightforward, in particular, at high densities when even bound electronic orbitals start to overlap \cite{Tilo_Nature_2023}. 

Density functional theory (DFT) has emerged as the de-facto workhorse of WDM theory \cite{wdm_book,vorberger2025roadmapwarmdensematter} and can also be utilized in a great variety of ways to interpret XRTS experiments \cite{Plagemann_NJP_2012, Gawne_ElectronicStructure_2025, Baczewski_PRL_2016, Dornheim_POP_2025, Witte_PRL_2017, Hentschel_POP_2023}. Vorberger \& Gericke \cite{Vorberger_2015} have demonstrated how elastic scattering can be described entirely by {\em ab inito} methods. 
Plagemann \textit{et al.} \cite{Plagemann_NJP_2012} have suggested to use the collision frequency obtained from a Kubo-Greenwood calculation for the optical conductivity to compute inelastic scattering at finite momentum transfer. This approach can be transferred to average atom models \cite{Johnson_PRE_2016} and has since been applied successfully for the analysis of a variety of XRTS experiments \cite{Witte_PRL_2017, Schoerner_PRE_2023}.

Recently, quasi-exact \emph{ab initio} path integral Monte Carlo (PIMC) simulations have been used to interpret XRTS experiments on strongly compressed beryllium \cite{Dornheim_NatComm_2025, Dornheim_POP_2025, schwalbe2025staticlineardensityresponse}. PIMC does not give direct access to $S_{ee}(q,\omega)$ itself, but it yields its two-sided Laplace transform, i.e., the intermediate scattering function in imaginary time, $F_{ee}(q,\tau)$ \cite{Dornheim_T_2022, Dornheim_T2_2022}. Although the numerical inversion of the Laplace transform to reconstruct $S(q,\omega)$ is notoriously difficult \cite{Jarrell_PhysRep_1996,Chuna_JPA_2025,dornheim_dynamic}, PIMC results can still be useful to compare with experimental observations, e.g., in the form of the Rayleigh weight $W_R(q)$~\cite{Dornheim_POP_2025} and the ratio of elastic to inelastic scattering~\cite{Dornheim_NatComm_2025}.

Approaches based on real-time time-dependent DFT \cite{Baczewski_PRL_2016, baczewski2021predictionsboundboundtransitionsignatures, White_ElectronicStructure_2025, kononov2025realtimetimedependentdensityfunctional} and linear-response time-dependent DFT \cite{Mo_PRL_2018, Moldabekov_PRR_2024, Moldabekov_MRE_2025, bespalov2026experimentalevidencebreakdownuniformelectrongas, moldabekov2025enhancingefficiencytimedependentdensity} have emerged as computationally expensive but practical ways to model XRTS over the last decade. In principle, these methods have the potential to become one of the most accurate tools for the calculation of XRTS spectra on a true \emph{ab initio} level.

The key feature that distinguishes these models and simulation techniques is whether they impose a clear-cut decomposition into bound and free states (chemical picture, Chihara) or treat the entire electron population holistically (physical picture, DFT and PIMC). In fact, the more sophisticated physical picture can be used to inform chemical models, see, e.g., the recent work by Bethkenhagen et al. \cite{PhysRevResearch.2.023260} on the charge state distribution of carbon and by Bellenbaum \textit{et al.}~\cite{Bellenbaum_PRR_2025} inferring ionization states and the depression of the ionization potential from PIMC simulations of warm dense hydrogen.

Most importantly, there have been several promising developments for the \emph{ab initio} description of the dynamic structure factor over a broad range of plasma parameters, and for a variety of materials. This includes the development of explicitly temperature-dependent exchange--correlation functionals for DFT \cite{ksdt, groth_prl, review, Karasiev_PRL_2018, karasiev_importance, kozlowski2023generalizedgradientapproximationthermal, Hilleke_PRM_2025,Karasiev_POP_2026}, the application of the Liouville-Lanczos method facilitating the computation of XRTS spectra at large momentum transfers in linear-response time-dependent DFT~\cite{Moldabekov_MRE_2025}, and new methods to deal with the notorious fermion sign problem in PIMC \cite{Xiong_JCP_2022, Dornheim_JCP_2023, Dornheim_NatComm_2025}.
In our opinion, the most difficult frontier will be the consistent treatment of non-equilibrium effects~\cite{Vorberger_PRE_2018}, which play a role in many situations with strongly driven plasmas being probed on very short time scales or directly by the heating x-ray beam \cite{Faustlin_PRL_2010, Chapman_PRL_2011, Chapman_HEDP_2012, Clerouin_PRE_2015, Fletcher_Frontiers_2022, Vorberger_PLA_2024, Bellenbaum_APL_2025}.

\subsection{Model-free analysis of XRTS experiments}
\label{sec:model_free}

In principle, measuring the dynamic structure factor yields direct access to many interesting physical information. Determining plasma parameters from the spectra without forward modeling is a very challenging but also very rewarding goal. For example, the detailed balance relation~\cite{quantum_theory}
\begin{eqnarray}
\label{eq:detailed_balance}
    S_{ee}(q,-\omega) = S_{ee}(q,\omega)\ e^{-\beta\hbar\omega}\ ,
\end{eqnarray}
with $\beta=1/k_\textnormal{B}T$ being the inverse temperature, is an exact relation in thermal equilibrium. It relates the up-shifted to the down-shifted side of $S(q,\omega)$ via a temperature-dependent factor. Accordingly, the ratio of the two sides determines the temperature~\cite{DOPPNER2009182}. Alas, it cannot be applied directly as the convolution Eq.~(\ref{eq:intensity}) destroys this useful symmetry.

Recently, Dornheim \textit{et al.} \cite{Dornheim_T_2022, Dornheim_T2_2022, Dornheim_MRE_2023} have suggested to switch from the usual frequency representation to the so-called imaginary-time domain.
In fact, $S_{ee}(q,\omega)$ is related to the imaginary time density--density correlation function (ITCF), which corresponds to the intermediate scattering function $F_{ee}(q,t)$ evaluated for an imaginary argument $t=-i\hbar\tau$ with $\tau\in[0,\beta]$, via a two-sided Laplace transform
\begin{eqnarray}\label{eq:Laplace}
    F_{ee}(q,\tau) = \mathcal{L}\left[S_{ee}(q,\omega)\right] = \int_{-\infty}^\infty \textnormal{d}\omega\ S_{ee}(q,\omega)\ e^{-\hbar\tau\omega}\ .
\end{eqnarray}
The well-known convolution theorem 
\begin{eqnarray}\label{eq:deconvolution}
   F_{ee}(q,\tau) = \mathcal{L}\left[S_{ee}\right] = \frac{\mathcal{L}\left[S_{ee}\circledast R\right]}{\mathcal{L}\left[R\right]}\ ,
\end{eqnarray}
where we have suppressed the $q$ and $\omega$ arguments for simplicity, then gives one direct access to the properly deconvolved ITCF.
We note that the transformation (\ref{eq:deconvolution}) is very stable with respect to the experimental noise, acting as a Laplace filter.
On the contrary, inverting the two-sided Laplace transform ($\mathcal{L}^{-1}[\dots]$) is exponentially ill-conditioned and, thus, generally not feasible for a set of experimental data~\cite{Chuna_JPA_2025}.
Nevertheless, $F_{ee}(q,\tau)$ contains, by definition, exactly the same information as $S_{ee}(q,\omega)$, only in an a-priori unfamiliar representation~\cite{Dornheim_MRE_2023,Dornheim_PTRS_2023}.
The ITCF on the left hand side of Eq.~(\ref{eq:deconvolution}) fulfills its own detailed balance relation in the imaginary time domain. Since its determination requires, in principle, only the measured data, that is the scattering intensity, the source spectrum and the characteristics of the detector, as inputs, one can use the detailed balance relation for the model-free extraction of the temperature from measured XRTS spectra \cite{Dornheim_T_2022,Dornheim_T2_2022}.

Subsequent to the temperature extraction method, the ITCF method has been further developed and now can provide the normalization of the spectra~\cite{Dornheim_SciRep_2024}, the Rayleigh weight~\cite{Dornheim_POP_2025}, and the static linear density response function~\cite{schwalbe2025staticlineardensityresponse}.
In addition, Vorberger \textit{et al.}~\cite{Vorberger_PLA_2024} suggested to use any deviation from the detailed balance in the imaginary-time domain to quantify non-equilibrium effects
and Gawne \textit{et al.}~\cite{gawne2025spectraldeconvolutiondeconvolutionextracting} showed how the need for any explicit knowledge of $R(\omega)$ can be eliminated in the ITCF analysis if simultaneous XRTS measurements at multiple scattering angles exist.
At the time of writing, the ITCF method has been applied to data sets that have been obtained at OMEGA~\cite{Dornheim_T_2022, Dornheim_T2_2022, Schoerner_PRE_2023}, LCLS~\cite{Dornheim_T_2022, Dornheim_T2_2022, Bellenbaum_APL_2025, bohme2026correlationfunctionmetrologywarm}, European XFEL~\cite{Dornheim_SciRep_2024, Smid_SciRep_2026}, NIF~\cite{Dornheim_NatComm_2025, Dornheim_SciRep_2024, schwalbe2025staticlineardensityresponse, Dornheim_POP_2025} and Shenguang-II laser facility~\cite{shi2025firstprinciplesanalysiswarmdense}.

\begin{figure*}[t!]\centering
\includegraphics[width=0.95\textwidth]{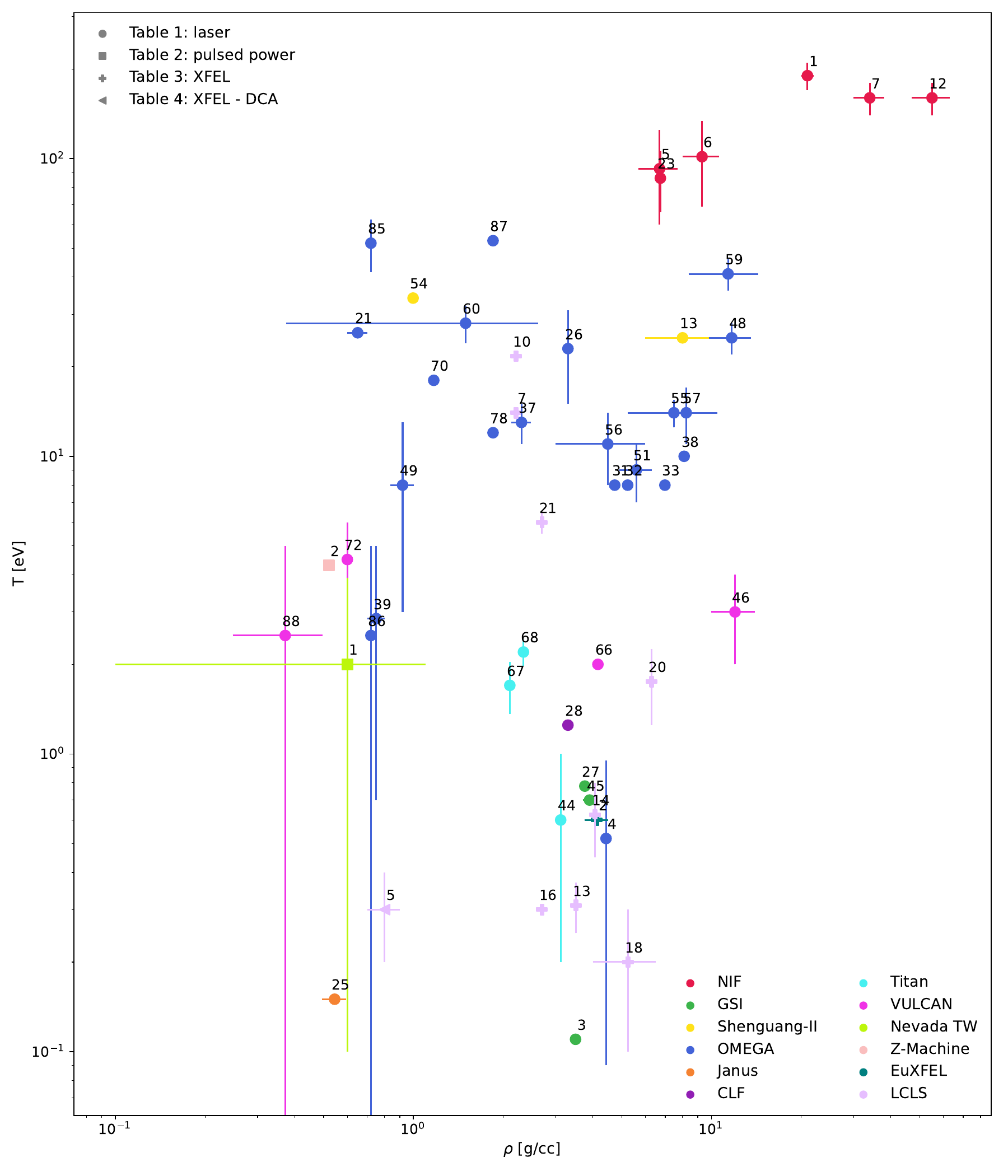}
\caption{\label{fig:fig} 
Overview of conditions that have been achieved in XRTS facilities in $\rho-T$ space.
This contains all datapoints where density and temperature were measured simultaneously, or the density could be obtained from electron number density and charge state estimates.
Different facilities are color-coded, the marker type indicates the Table and facility type.
Numbers indicate the entry in a given table.
Errorbars are included where present in the original work.
}
\end{figure*}

\section{Overview of XRTS experiments}
\label{sec:overview}

In the following, we give an extensive overview of over 90 different XRTS experiments, grouped on the highest level by the set-up and the type of facility where the experiment has been carried out:
Tab.~\ref{table:laser} contains experiments at optical laser facilities (OMEGA, NIF, GSI, Rutherford Appleton, Shenguang-II, LULI-2000, as well as Janus, Titan and Jupiter in Livermore);
in Tab.~\ref{table:pulsed}, XRTS experiments at pulsed power facilities (Nevada TW and Sandia Z-machine) are listed;
Tabs.~\ref{table:XFEL} and \ref{table:DCA} present experiments at x-ray free electron lasers (FLASH, LCLS and European XFEL) with resolutions both in the eV and meV range, i.e., campaigns without and with the high resolution provided by diced crystal analyzers, respectively; 
finally, Tab.~\ref{table:synchrotron} contains a selection of a few experiments performed at synchrotron facilities and storage rings (DESY, Diamond Light Source UK, ESRF).
For some experiments, we have also included subsequent re-evaluations based on other methods. 
We note that we have only included experiments with related publications that actually show the measured XRTS spectra either in the main text, or in the supplemental material.

Fig.~\ref{fig:fig} contains an overview of the experiments described in detail below.
Here, we show measurements at different facilities in $\rho-T$ space for datapoints where either density and temperature where measured simultaneously, or the mass density could be extracted from inferred number density and charge state.
Errorbars were included where available in the original publication.

We also want to highlight some special insights on the properties of dense
matter that have been enabled by XRTS. These range from the determination of
standard thermodynamic properties and charge states in dense plasmas to the 
investigation of transport and relaxation properties and the collective modes 
in materials with strongly coupled particles. In the following summary, we will 
roughly follow the time-line of the investigations but bundle similar physics. 

The first application of XRTS was the determination of the basic plasma 
properties similar to optical Thomson scattering in low-density plasmas
\cite{sheffield2010plasma, Glenzer_1999}. First, Landen et al.\ 
\cite{LANDEN2001465} published the
theoretical framework. This was quickly followed by the first experimental 
results \cite{Gregori_AIP_2002} and then the landmark study by 
Glenzer et al.\ that demonstrated temperature and charge state measurements in 
isochorically heated Beryllium \cite{Glenzer_PRL_2003}. Probing a material 
with fixed density proved very helpful for the analysis as only the 
two parameters needed to be fit to theoretical predictions. Later, this 
restriction was lifted when shock-compressed beryllium was diagnosed
\cite{Lee_PRL_2009}. With further improvements over the next decade, XRTS was
made so versatile that it allowed for the probing of states in-flight during 
the compression of a spherically driven capsule 
\cite{Kritcher_PRL_2011, Poole_PoP_2022}, which has
direct application to inertial confinement fusion. One common feature of the
first experiments was the use of a back-scattering geometry that mainly probed 
the single particle behavior and, thus, revealed properties of the 
distribution function \cite{sheffield2010plasma} including matter with 
degenerate electrons.

After the determination of the plasma parameters was established, investigations
of the ion structure in warm dense matter became possible by using the elastic
Rayleigh feature of the XRTS spectra.
Important steps were the push to smaller scattering angles, i.e., momentum
transfers, \cite{GarciaSaiz2008} and multiple $q$-points \cite{Ma_PRL_2013}
to fully benchmark 
theoretical predictions that deviated significantly even for light elements.
For larger ions, an additional repulsion due to the interaction of full
bound shells was predicted \cite{Wunsch_PRE_2009} and later also found
experimentally \cite{Ma_PRL_2013,Ma_PoP_2014} in aluminum at warm dense matter
conditions. These full electron shell can also influence the screening 
\cite{Gericke_PRE_2010} and XRTS experiments were able to verify this effect as well 
\cite{Fletcher2015}. These investigations continue and now focus on elements
with a more complex ion structure like carbon \cite{Kraus_Nature_2025}. However, it is
interesting to notice that even for beryllium at normal density, the long 
wavelength behavior of the elastic scattering feature is not sufficiently
explained to date.

The next major step toward a full use of XRTS was done by experiments focusing
on the inelastic scattering at small scattering angles. In this way, the
collective behavior of the electrons, i.e.\ plasmons, can be investigated.
Since the plasmons are mainly damped by electron-ion interactions, this
feature also reveals the collisionality in the system supplementing other
optical measurements \cite{kremp_book}. The first plasmon feature was resolved with 
an optically driven x-ray source on solid-density beryllium at the Omega 
facility \cite{Glenzer_PRL_2007}. Although clearly demonstrating plasmon 
scattering, it and following experiments \cite{Neumayer_PRL_2010} showed that 
this approach suffers from large noise levels due to the weak scattering and 
relatively low photon numbers. The latter point can be overcome at free x-ray 
laser facilities that, indeed, showed their great potential for plasmon 
scattering \cite{Sperling_PRL_2015, Witte_PRL_2017, bespalov2026experimentalevidencebreakdownuniformelectrongas}, especially when the x-ray
beam uses seeded beams \cite{Fletcher2015}. 

XFELS enable also many sequential shots and, thus, averaging can strongly 
reduce the noise level of the data \cite{Fletcher_RSI_2014}. Combined with the 
seeded beam, such averaging allowed to fully resolve the plasmon feature in
cold aluminium \cite{Fletcher2015}. Driven targets were however still much more 
restricted due to target delivery and thus showed much less accuracy. Further 
developments on gas jets could overcome this by continuously replenishing the 
targets \cite{Faustlin_PRL_2010,Zastrau,Toleikis_2010}. 
Another possibility XFELs allow is the heating of the target by the x-rays 
themselves. Such self-heating experiments \cite{Faustlin_PRL_2010} allow for 
very unique states to be reached in a x-ray-only setup 
\cite{Sperling_PRL_2015,Witte_PRL_2017,kraus_xrts,Bellenbaum_APL_2025} (see also the recent TDDFT predictions by Moldabekov \textit{et al.}~\cite{Moldabekov_PRR_2024,Moldabekov_ACS_OMEGA_2024}) but are 
usually fully dynamic. This aspect can be problematic for the analysis but also 
opens the door for investigating the nonequilibrium behavior of dense matter.

In the last decade, an exciting new avenue for the application of XRTS has 
opened: the diagnostics of states with extremely high densities created by 
spherical implosions at the NIF. These experiments require extremely  
hard x-rays to penetrate the dense samples and precise timing
to catch the states with the highest densities in highly dynamic targets
\cite{Doeppner_RSI_2016}.
The first experiments \cite{Kraus_PRE_2016} used solid spheres and showed that 
often-used models for the ionisation degree fail at these extreme conditions. 
Using shell targets, the created densities can be further increased 
significantly by the shock wave bouncing of central symmetry point 
\cite{Bishel_RSI_2018}. With this technique the Gbar-range could be reached and 
fully diagnosed on the atomic level \cite{Tilo_Nature_2023}, which revealed 
strong modifications for the bound states at these densities. Other properties, 
like the opacity, can also be investigated with this setup \cite{Lutgert_PoP_2022}. 
With these experimental techniques, the conditions in the deep interior of 
giant planets and the outer layers of small stars are now in reach of laboratory 
investigations. 

The second exciting class of investigation that was enabled recently is the
observation of ion acoustic modes, that is, the analog to the plasmon
feature seen in the inelastic scattering signal. For these experiments, the
scattering signal must be resolved with a very high resolution of a few meV
\cite{Gregori_PoP_2009} but they offer new insights into the physics of 
dense matter \cite{Vorberger_PRL_2012, Maybey_NatComm_2017}. This resolution 
can neither be reached with
laser-created x-ray sources nor XFEL pulses directly. It was however made 
possible by the developments of a 4-bounce silicon monochromator and diced 
crystal analyzers \cite{VERBENI_2005, Descamps_2022}. With these advances, 
the mode structure related to the fluctuations of the nuclei could be observed
\cite{McBride_RSI_2018,Wollenweber_RSI_2021} with an approach
that was tested on solid samples with well-known phonon structures
\cite{Descamps_SciReports_2020,Gawne_PRB_2024}. The analysis of this acoustic
mode structure allowed to determine the sound speed in methane 
\cite{White_PRR_2024}. Recently, this high-resolution technique has been 
applied in back-scattering geometry where is reveals the one-particle dynamics 
of the nuclei. With this setup, strongly overheated states in gold 
\cite{White_Nature_2025} could be observed. 

A sign for the maturity of XRTS as a diagnostic method is the fact that it is
now used as a secondary diagnostics establishing plasma conditions. In this
way, the primary property investigated can be related to a well-defined state
making comparisons with theoretical predictions much more insightful. 
Examples are measurements of electron-ion energy exchange rates
\cite{Fletcher_Frontiers_2022} and transport properties like the stopping power
\cite{Lahmann_PPCF_2023} or the sound speed \cite{White_PRR_2024}. Moreover, 
XRTS was used to support 
techniques like x-ray photon correlation spectroscopy \cite{Heaton_PRR_2025}
and might help to resolve the discrepancies in the x-ray transmitivity of
ICF ablators \cite{Hall_PoP_2024} with precise measurements of the ion charge state 
in highly compressed matter.

\begin{widetext}
\onecolumngrid
\begingroup
\small
\begin{longtblr}[
  caption = {Overview of XRTS experiments at laser facilities ordered by year of publication.},
  label = {table:laser}
]{colspec = {Q[0.1,l] Q[1.5,l] Q[1.5,l] Q[1.2,l] Q[1.2,l] Q[3.0,l] Q[3.2,l]  Q[3.0,l]},
  rowhead = 1,
  columns={font=\footnotesize} } 
    & Material & Facility & $E_0$ & $\theta$ & conditions & method & reference \\
    \hline
    \hline
    1 & Be (capsule) implosion & NIF & $9\,$keV   & $75^\circ$    & $T=190\pm20\,$eV, $\rho=21\pm1\,$g/cc  & ITCF/PIMC   elastic/inelastic    & Dornheim \textit{et al.}~\cite{Dornheim_NatComm_2025} (2025) \\\hline
    2 & diamond & GSI (PHELIX) & $4.75\,$ keV   & $135^\circ$  & ambient  & ---   & Hesselbach \textit{et al.}~\cite{Hesselbach_MRE_2024} (2024)\\\hline
    3 & diamond, ion heating & GSI (PHELIX) & $4.75\,$keV   & $135^\circ$  & $T\approx0.11\,$eV, $\rho=\rho_0$  & DFT-MD / elastic-to-inelastic   & L\"utgert \textit{et al.}~\cite{Luetgert_MRE_2024} (2024)\\\hline
    4 & Si (foil) shocked & OMEGA & $8.4\,$keV   & 70$^\circ$, 95$^\circ$, 98$^\circ$   & $0.09\,$eV$<T<0.95\,$eV  $\rho=4.43\pm0.08\,$g/cc  & Chihara   Rankine-Hugoniot   & Poole \textit{et al.}~\cite{Poole_PRR_2024} (2024)\\\hline
    5 & \SetCell[r=8]{h}Be (capsule) implosion & NIF & $9\,$keV   & $120^\circ$    &  (a) $T=60-125\,$eV, $\rho=6.7\pm 1\,$g/cc $Z=3.0\pm0.1$ & Chihara     & D\"oppner \textit{et al.}~\cite{Tilo_Nature_2023} (2023) \\\cline{5-7}
    6 & &  &  &  & (b) $T=69-134\,$eV, $\rho=9.3\pm1.3\,$g/cc  $Z=3.05\pm0.1$ & Chihara     & D\"oppner \textit{et al.}~\cite{Tilo_Nature_2023} (2023) \\ \cline{5-7}
    7 & &  &  &  & (c) $T=160\pm20\,$eV, $\rho=34\pm4\,$g/cc $Z=3.4\pm0.1$ & Chihara     & D\"oppner \textit{et al.}~\cite{Tilo_Nature_2023} (2023)  \\ \cline[dotted]{5-7}
    8 & &  &  &  & $T=155\pm15\,$eV, $\rho=22\pm2\,$g/cc, $Z=3.6\pm0.1$  & ITCF/PIMC    elastic/inelastic  & {Dornheim \textit{et al.}~\cite{Dornheim_NatComm_2025} (2025)} \\\cline[dotted]{5-7}
    9 & &  &  &  & $T=155\pm15\,$eV, $\rho=22\pm2\,$g/cc, $Z=3.5\pm0.1$  & DFT-MD  Rayleigh weight   & Dornheim \textit{et al.}~\cite{Dornheim_POP_2025} (2025) \\\cline[dotted]{5-7}
    10 & &  &  &  & $T=155\pm15\,$eV, $\rho=18\pm6\,$g/cc  & ITCF/PIMC   density response  & Schwalbe \textit{et al.}~\cite{schwalbe2025staticlineardensityresponse} (2025) \\\cline[dotted]{5-7}
    11 & &  &  &  & $T=155\pm5\,$eV, $\rho=20\pm2\,$g/cc, $Z=3.25\pm0.01$  & neutral pseudoatom   Rayleigh weight  & Dharma-wardana and Klug~\cite{Dharma_wardana_PRE_2025} (2025) \\\cline{5-7}
    12 & &  &  &  & (d) $T=160\pm20\,$eV, $\rho=55\pm8\,$g/cc,  $Z=3.3\pm0.1$  & Chihara     & D\"oppner \textit{et al.}~\cite{Tilo_Nature_2023} (2023) \\ \hline
    13 & C-H-Cl (pellet), double-cone experiment & Shen-guang-II & $8.95\,$keV   &  $120^\circ$    &  $T=25\,$eV, $\rho=8\pm2\,$g/cc  & ITCF, DFT   & Shi \textit{et al.}~\cite{15-20220361,shi2025firstprinciplesanalysiswarmdense} (2022/25)  \\ \hline
    14 & Graphite x-ray heated & Shen-guang-II & $4.75\,$keV   &  $90^\circ$    &  $T=4.4\pm1.2\,$eV,  $n_e=1.24-2.24\times10^{23}$g/cc  & Chihara     & Lv \textit{et al.}~\cite{Lv_POP_2019} (2019)  \\\hline
    15 & \SetCell[r=4]{h} CH (foam) laser driven blast wave & OMEGA & $7.78\,$keV  & $100\pm5^\circ$ & (a) $T_e=7\,$eV, $Z_C=2$, $Z_H=0.5$ & Chihara   & Falk \textit{et al.}~\cite{Falk_PRL_2018,Falk_PPCF_2020} (2018/20) \\ \cline{5-7}
    16 &  &  &  &  & (b) $T_e=17.5\pm2.5\,$eV, $Z=2.5$  & Chihara   & Falk \textit{et al.}~\cite{Falk_PRL_2018,Falk_PPCF_2020} (2018/20) \\ \cline{5-7}
    17 &  &  &  &  & (c) $T_e\approx25\pm5\,$eV, $Z=2.2-2.9$  & Chihara   & Falk \textit{et al.}~\cite{Falk_PRL_2018,Falk_PPCF_2020} (2018/20)  \\\cline{5-7}
    18 &  &  &  &  & (d) $T_e=35\pm5$, $Z=2.2$ & Chihara   & Falk \textit{et al.}~\cite{Falk_PRL_2018,Falk_PPCF_2020} (2018/20) \\\hline
    19 & Be (cylinder), isochoric heating & OMEGA & $9\,$keV  & $120^\circ$    & $T=5\substack{+5 \\ -5}$\,eV, $T_{Be}<2.2$, $\rho\approx\rho_0$ & Chihara    & Saunders \textit{et al.}~\cite{Saunders_PRE_2018,Lahmann_PPCF_2023} (2018) \\\hline
    20 & B (cylinder), isochoric heating & OMEGA & $9\,$keV  & $120^\circ$   & $T=8.6\substack{+2.7 \\ -3.1}$\,eV, $T_{B}<3.1$, $\rho\approx\rho_0$ & Chihara    & Saunders \textit{et al.}~\cite{Saunders_PRE_2018,Lahmann_PPCF_2023} (2018) \\\hline
    21 & CH (foam) shocked released & OMEGA & $7.8\,$keV   & $100\pm5^\circ$    & $T=26\,$eV, $\rho=0.6-0.7\,$g/cc & Chihara     & Falk \textit{et al.}~\cite{Falk_PPCF_2017} (2017) \\\hline
    22 & Graphite  shocked released & OMEGA & $7.8\,$keV   & $100\pm5^\circ$    & $T=12\,$eV  & Chihara     & Falk \textit{et al.}~\cite{Falk_PPCF_2017} (2017) \\\hline
    23 & CH (capsule) implosion & NIF & $9\,$keV   & $84^\circ$    & $\rho=6.74\,$g/cc, $T=86\pm20\,$eV, $Z_C=4.92\pm0.15$  & Chihara, hydro ($\rho$)     & Kraus \textit{et al.}~\cite{Kraus_PRE_2016} (2016) \\\hline
    24 & RF (C,H,O) (foam) shocked & OMEGA & $7.8\,$keV   & $90\pm20^\circ$   & $T\lesssim20\,$eV, $Z_C=Z_O\lesssim2$ &  Chihara    & Belancourt \textit{et al.}~\cite{Belancourt_RSI_2016} (2016) \\\hline
    25 & Deuterium shocked & Janus (LLNL) & $2\,$keV   & $35^\circ$, $135^\circ$   & $T_e=0.15\,$eV, $\rho=3.2\substack{+0.2 \\ -0.3}\rho_0$ ($\rho_0=0.17\pm0.004\,$g/cc), $Z=0.15\pm0.08$ &  Chihara / DFT-MD    & Davis \textit{et al.}~\cite{Davis_NatComm_2016} (2016) \\\hline
    26 & CH$_2$ (capsule), hohlraum & OMEGA & $9\,$keV   & $87\pm15^\circ$   & $T_e=23\substack{+7 \\ -8}\,$eV, $\rho=3.3\,$g/cc, $Z_H=1$, $Z_C=4$ &  Chihara ($T_e$), rad-hydro ($\rho$)    & Saunders \textit{et al.}~\cite{Saunders_RSI_2016} (2016) \\\hline
    27 & Graphite, shocked & GSI (PHELIX) & $4.75\,$keV   & $126^\circ$    & $T=0.78\,$eV $\rho=3.76\,$/cc   & MULTI-2D    & Kraus \textit{et al.}~\cite{Kraus_POP_2015} (2015) \\\hline
    28 & Graphite, shocked & CLF, Rutherford Appleton & $4.95\,$keV   & $126^\circ$    & $T=1.25\,$eV $\rho=3.30\,$/cc   & HELIOS-1D, Chihara    & Kraus/Helfrich \textit{et al.}~\cite{Kraus_POP_2015,Helfrich_HEDP_2015} (2015) \\\hline
    29 & CH (capsule) shocks & OMEGA & $9\,$keV   & $135\pm15^\circ$    & $T=10.5\substack{+6.5 \\ -0}$\,eV $n_e=1.31\pm0.37\times10^{24}\,$/cc   & Chihara / Rayleigh weight    & Chapman \textit{et al.}~\cite{Chapman_NatComm_2015} (2015) \\ \hline
    30 & \SetCell[r=4]{h} Carbon (capsule) implosion & OMEGA & $9\,$keV   & $135\pm10^\circ$    & (a) $T=6\,$eV, $n_e=0.80\times10^{24}$/cc & Chihara     & Fletcher \textit{et al.}~\cite{Fletcher_PRL_2014} (2014) \\ \cline{5-7}
    31 & &  &  &  & (b) $T=8\,$eV, $Z=4$, $n_e=0.95\times10^{24}$/cc & Chihara     & Fletcher \textit{et al.}~\cite{Fletcher_PRL_2014} (2014) \\\cline{5-7}
    32 & &  &  &  & (c) $T=8\,$eV, $Z=4$, $n_e=1.05\times10^{24}$/cc & Chihara     & Fletcher \textit{et al.}~\cite{Fletcher_PRL_2014} (2014) \\\cline{5-7}
    33 & &  &  &  & (d) $T=10\,$eV, $Z=4$, $n_e=1.4\times10^{24}$/cc & Chihara     & Fletcher \textit{et al.}~\cite{Fletcher_PRL_2014} (2014) \\ \hline
    34 & \SetCell[r=2]{h} Carbon (foam) shocked & OMEGA & $7.8\,$keV   & $90\pm8^\circ$    & (a) $T=25\pm5\,$eV, $Z=2.3\pm1$, $3-3.5\times\rho_0$   & Chihara     & Gamboa \textit{et al.}~\cite{Gamboa_POP_2014,Gamboa_HEDP_2014} (2014) \\\cline{5-7}
    35 & &  &  &  & (b) all data sets~\cite{Gamboa_HEDP_2014}: $T=20-40\,$eV, $Z=2-4$   & Chihara     & Gamboa \textit{et al.}~\cite{Gamboa_POP_2014,Gamboa_HEDP_2014} (2014) \\\hline
    36 & CH (capsule) implosion & OMEGA & $9\,$keV   & $135\pm15^\circ$    & $T_e=26\pm3\,$eV, $Z_C\approx2.5$, $n_e=2.4\times10^{24}$/cc  & Chihara     & Kritcher \textit{et al.}~\cite{KRITCHER201427} (2014)\\\hline
    37 & Diamond shocked released & OMEGA & $7.8\,$keV   & $100\pm5^\circ$    & $T=13\pm2\,$eV, $Z=2.25\pm0.15$, $n_e=2.6\pm0.2\times10^{23}$/cc  & Chihara     & Falk \textit{et al.}~\cite{Falk_PRL_2014,Falk_POP_2014} (2014) \\\hline
    38 & Al (foil) shocked & OMEGA & $17.9\,$keV   & $69^\circ$, $111^\circ$    & $\rho=8.1\,$g/cc, $T=10\,$eV, $Z=3$ & Chihara, Rayleigh weight     & Ma \textit{et al.}~\cite{Ma_PRL_2013,Ma_PoP_2014} (2013/14); see also average-atom calculations by Souza \textit{et al.}~\cite{Souza_PRE_2014} and Hou \textit{et al.}~\cite{Hou_PRE_2015}\\\hline
    39 & Deuterium planar shock& OMEGA & $2.96\,$keV   & $90^\circ$    & $T_e=0.7-5\,$eV, $Z=0.6-1$, $\rho=0.7-0.8\,$g/cc & Chihara     & Falk \textit{et al.}~\cite{Falk_PRE_2013} (2013)\\\hline
    40 & \SetCell[r=4]{h} CH (capsule) implosion & OMEGA & $9\,$keV   & $135^\circ$    & (a) $T_e=6\,$eV, $Z_C=4$, $n_e=0.8\times10^{24}$/cc  & Chihara     & Fletcher \textit{et al.}~\cite{Fletcher_POP_2013} (2013)\\ \cline{5-7}
    41 & &  &  &  & (b) $T_e=6\,$eV, $Z_C=4$, $n_e=0.95\times10^{24}$/cc  & Chihara     & Fletcher \textit{et al.}~\cite{Fletcher_POP_2013}  (2013)\\ \cline{5-7}
    42 & &  &  &  & (c) $T_e=8\,$eV, $Z_C=4$, $n_e=1.05\times10^{24}$/cc  & Chihara     & Fletcher \textit{et al.}~\cite{Fletcher_POP_2013}  (2013) \\ \cline{5-7}
    43 & &  &  &  & (d) $T_e=10\,$eV, $Z_C=4$, $n_e=1.4\times10^{24}$/cc  & Chihara     & Fletcher \textit{et al.}~\cite{Fletcher_POP_2013} (2013)\\ \hline
    44 &  B (solid) planar shocked & Titan LLNL & $4.95\,$keV & $31^\circ$, $47^\circ$,$62^\circ$, $67^\circ$   & $T=0.2-1\,$eV, $\rho=1.5\rho_0$,$n_e=4\times10^{23}$/cc   & Chihara     & LePape \textit{et al.}~\cite{LePape_NJP_2013} (2013) \\ \hline
    45 &  Graphite shocked & GSI (PHELIX) & $4.75\,$keV    & $105^\circ$, $126^\circ$   &  $\rho=3.9\pm0.2\,$g/cc, $T\approx0.7\,$eV   & Chihara / DFT-MD     & Kraus \textit{et al.}~\cite{Kraus_PRL_2013} (2013) \\ \hline
    46 &  Fe (foil) shocks & VULCAN Rutherford Appleton & $4.75\,$keV & $38\pm7^\circ$, $58\pm7^\circ$, $90\pm7^\circ$  & $10\,$g/cc$<\rho<14\,$g/cc, $2\,$eV$<T<4\,$eV   & rad hydro     & White \textit{et al.}~\cite{White_HEDP_2013} (2013) \\ \hline
    47 & Ar (gas) shocked & OMEGA & $6.2\,$keV   & $79\pm16^\circ$   &  $T=34\,$eV $\pm14\%$ & Chihara     & Visco \textit{et al.}~\cite{Visco_PRL_2012} (2012) \\ \hline
    48 & Be (foil) colliding shocks & OMEGA & $9\,$keV   & $140\pm10^\circ$   & $T=25\pm3\,$eV, $\rho=11.7\pm1.9$g/cc, $n_e=1.8\pm0.2\times10^{24}$/cc & Chihara     & Fortmann \textit{et al.}~\cite{Fortmann_PRL_2012} (2012) \\ \hline
    49 & Deuterium (liquid) shocked & OMEGA & $2.96\,$keV   & $87.8^\circ$   & $T_e=8\pm5\,$eV, $n_e=2.2\pm0.5 \times10 ^{23}$/cc, $Z=0.8\substack{+0.15 \\ -0.25}$\,eV & Chihara     & Regan \textit{et al.}~\cite{Regan_PRL_2012} (2012) \\ \hline
    50 & Deuterium (liquid) cryogenic & Janus (LLNL) & $2.005\,$keV   & $45^\circ$   & ambient & ---     & Davis \textit{et al.}~\cite{Davis_Instrumentation_2012} (2012) \\ \hline
    51 & Be (foil) single shocked & OMEGA & $9\,$keV   & $140\pm10^\circ$   & $T=9\pm2\,$eV, $\rho=5.6\pm0.7$g/cc, $n_e=8.6\pm0.7\times10^{23}$/cc & Chihara     & Fortmann \textit{et al.}~\cite{Fortmann_PRL_2012} (2012)\\ \hline
    52 &  Carbon shocked & GSI (PHELIX) & $4.75\,$keV   & $126\pm10^\circ$    & $\rho=3.9\pm0.4\,$g/cc  &  Rankine-Hugoniot     & Kraus \textit{et al.}~\cite{Kraus_HEDP_2012} (2012) \\ \hline
    53 & C$_6$H$_{12}$ (foam) x-ray heated& Sheng-guang-II & $4.75\,$keV   & $90^\circ$    & $T=5\,$eV, $1\,$g/cc, $Z_H=Z_C=1$  & Chihara     & Hu \textit{et al.}~\cite{Hu_PST_2012} (2012) \\ \hline
    54 & Carbon (foam) x-ray heated & Sheng-guang-II & $4.75\,$keV   & $66.5^\circ$    & $T=34\,$eV, $1\,$g/cc, $n_e=1.6\times10^{23}$, $Z=3.2$  & Chihara     & Bao \textit{et al.}~\cite{Bao_POP_2012} (2012)\\ \hline
    55 & \SetCell[r=2]{h} Be (capsule) implosion & OMEGA & $9\,$keV   & $113^\circ$    & (a) $T=14\pm1.5\,$eV, $Z=2$, $n_e=1.0\pm0.3\times10^{24}$/cc  & Chihara     & Kritcher \textit{et al.}~\cite{KRITCHER2011271} (2011)\\ \cline{5-7}
    56 &  &  &  &  & (b) $T=11\pm3\,$eV, $Z=2$, $n_e=6.0\pm2\times10^{23}$/cc  & Chihara     & Kritcher \textit{et al.}~\cite{KRITCHER2011271} (2011) \\ \hline
    57 & \SetCell[r=4]{h} Be (capsule) implosion & OMEGA & $9\,$keV   & $135\pm15^\circ$    & (a) $T=14\pm3\,$eV, $Z\approx2$, $n_e=1.1\pm0.3\times10^{24}$/cc  & Chihara     &  Kritcher \textit{et al.}~\cite{Kritcher_PRL_2011,Kritcher_AIP_2012} (2011)\\ \cline[dotted]{5-7}
    58 & &  &  &  & $\rho=7.9\pm0.9\,$g/cc  & LR-TDDFT     & Sch\"orner \textit{et al.}~\cite{Schoerner_PRE_2023} (2023)\\ \cline{5-7}
    59 & &  &  &  & (b) $T=41\pm5\,$eV, $Z\approx2.5$, $n_e=1.9\pm0.5\times10^{24}$g/cc  & Chihara     & Kritcher \textit{et al.}~\cite{Kritcher_PRL_2011,Kritcher_AIP_2012} (2011)\\ \cline{5-7}
    60 & &  &  &  & (c) $T=28\pm4\,$eV, $Z\approx2$, $n_e=0.2\pm0.15\times10^{24}$/cc  & Chihara     & Kritcher \textit{et al.}~\cite{Kritcher_PRL_2011,Kritcher_AIP_2012} (2011)\\ \hline
    61 & B (foil) shocked compressed & Jupiter LLNL & $4.95\,$keV   & $31-67^\circ$ ($\pm10\%$)   & $T=0.2\,$eV, $n_e=4\times10^{23}\,$cc  & Born-Mermin  (+LFC~\cite{farid}) & Neumayer \textit{et al.}~\cite{Neumayer_PRL_2010} (2010)\\\hline
    62 & B (foil) shocked & Titan LLNL & $4.95\,$keV, $8.05\,$keV     & $31^\circ$, $63^\circ$   & $n_e=4\times10^{23}\pm18\%\,$/cc, $\rho=3.15\pm0.7\,$g/cc, low $T$   & Chihara / Bohm-Gross    & LePape \textit{et al.}~\cite{LePape_POP_2010} (2010) \\ \hline
    63 & \SetCell[r=2]{h} Be (foil) shock& OMEGA & $6.2\,$keV   & $90^\circ$   & (a) $T=13\,$eV, $n_e=7.5\times10^{23}\,$cc ($\pm7\%$)  & Chihara   & Lee \textit{et al.}~\cite{Lee_PRL_2009} (2009); see also average-atom calculations by Souza \textit{et al.}~\cite{Souza_PRE_2014}\\ \cline[dotted]{5-7}
    64 & &  &  &  & $T=15\substack{+21 \\ -14}$\,eV, $\rho=5\substack{+8 \\ -4}$\,g/cc, $Z=1.8\substack{+0.5 \\ -1}$  & Chihara+MCMC     & Kasim \textit{et al.}~\cite{Kasim_POP_2019} (2019)\\ \cline{4-7}
    65 & &  &  &  $25^\circ$   & (b) $T=13\pm3\,$eV, $n_e=7.5\times10^{23}\,$g/cc ($\pm6\%$)  & Chihara     & Lee \textit{et al.}~\cite{Lee_PRL_2009} (2009); see also average-atom calculations by Souza \textit{et al.}~\cite{Souza_PRE_2014} and DFT-MD simulations by Plagemann \textit{et al.}~\cite{Plagemann_PRE_2015}\\ \hline
    66 & Li (foil) shocked& VULCAN & $2.96\,$keV   &  $120^\circ$   & $T_e\approx T_i  \approx 2\,$eV, $\rho\approx 2\rho_0$  & Chihara     & Kugland \textit{et al.}~\cite{Kugland_PRE_2009} (2009)\\ \hline
    67 & \SetCell[r=2]{h} LiH (foil) shock& Titan LLNL & $4.51\,$keV   &   $95^\circ$   & (a) $T=1.7\,$eV ($\pm20\%$), $\rho=2.7\rho_0$, $n_e=1.6\times10^{23}$/cc ($\pm20\%$) & Chihara     & Kritcher \textit{et al.}~\cite{Kritcher_PRL_2009} (2009) \\ \cline{5-7}
    68 &  &  &  &  & (b) $T=2.2\,$eV ($\pm10\%$), $\rho=3\rho_0$,$n_e=1.7\times10^{23}$/cc ($\pm21\%$)  & Chihara     & Kritcher \textit{et al.}~\cite{Kritcher_Science_2008,Kritcher_POP_2009} (2008/09)\\ \hline
    69 & LiH (foil) shock& Titan LLNL& $4.51\,$keV   &  $40\pm10^\circ$   &  $T\leq0.4\,$eV, $Z\leq0.2$ &  Chihara     & Kritcher \textit{et al.}~\cite{Kritcher_Science_2008,Kritcher_POP_2009} (2008/09)\\ \hline
    70 & \SetCell[r=2]{h} Be (solid) x-ray heated & OMEGA & $2.96\,$keV   & $40^\circ$    & $T=18\,$eV, $\rho=1.17\,$g/cc, $n_e=1.8\times10^{23}$/cc, $Z\approx2.3$  & RPA fit  (detailed balance)   & D\"oppner \textit{et al.}~\cite{DOPPNER2009182,Dopp} (2009); see also the average-atom calculation by Johnson \textit{et al.}~\cite{Johnson_PRE_2012} \\ \cline[dotted]{5-7}
    71 & &  &  &  & $T=19\pm1.5\,$eV, $\rho=1.8\,$g/cc, $Z\approx2.14$  & ITCF/ LR-TDDFT  (TRK sum rule)   & Sch\"orner \textit{et al.}~\cite{Schoerner_PRE_2023} (2023) \\ \hline
    72 & Li (foil) shock& Vulcan Rutherford Appleton & $2.96\,$keV   & $60^\circ$, $40^\circ$    & $T=4.5\pm1.5\,$eV, $Z=1.35\pm0.1$, $\rho=0.6\pm0.025\,$g/cc  & Chihara     & Garcia Saiz \textit{et al.}~\cite{GarciaSaiz2008} (2008) \\\hline
    73 & \SetCell[r=3]{h}Carbon foam x-ray heated & OMEGA & $4.75\,$keV   & $95^\circ$    & (a) $T_e=70\,$eV, $Z_C=4.6$  & Chihara     & Gregori \textit{et al.}~\cite{Gregori_PRL_2008} (2008) \\\cline{5-7}
    74 & &  &  &  & (b) $T_e=130\,$eV, $Z_C=5.2$ & Chihara     & Gregori \textit{et al.}~\cite{Gregori_PRL_2008} (2008) \\\cline{5-7}
    75 & &  &  &  & (c) $T_e=60\,$eV, $Z_C=4.6$  & Chihara     & Gregori \textit{et al.}~\cite{Gregori_PRL_2008} (2008) \\ \hline
    76 & CH foil, shocked  & OMEGA & $9\,$keV   & $120\pm10^\circ$, $90\pm10^\circ$    & $T\approx10\,$eV, $\rho\approx4\rho_0$ & Chihara     & Sawada \textit{et al.}~\cite{Sawada_POP_2007} (2007)\\ \hline
    77 & CH/Al/CH multi-layer, shocked  & LULI-2000 & $4.7\,$keV   & $47^\circ$    & $T_{Al}\sim0.35\,$eV, $T_{CH}\sim0.8\,$eV, $n_e\sim2.5\times10^{23}$/cc & Chihara + Debye-Waller factor & Ravasio \textit{et al.}~\cite{Ravasio_PRL_2007} (2007)\\\hline
    78 & \SetCell[r=2]{h} Be (solid) x-ray heated & OMEGA & $2.96\,$keV   & $40^\circ$    & $T\approx12\,$eV, $\rho=1.85\,$g/cc  & Chihara\newline Mermin (RPA)     & Glenzer \textit{et al.}~\cite{Glenzer_PRL_2007} (2007); see also non-equilibrium analysis by Chapman and Gericke~\cite{Chapman_PRL_2011} and DFT-MD simulations by Plagemann \textit{et al.}~\cite{Plagemann_PRE_2015} \\ \cline[dotted]{5-7}
    79 & &  &  &  & $T=14.8\pm2\,$eV  &   ITCF     & Dornheim \textit{et al.}~\cite{Dornheim_T_2022,Dornheim_T2_2022} (2022/23) \\ \hline
    80 & \SetCell[r=4]{h} Hydrocarbon (gas bag) laser heated & OMEGA & $9\,$keV   & $120\pm5^\circ$   &   (a) $T_e=280\,eV$ & Chihara     & Gregori \textit{et al.}~\cite{Gregori_JQuantSpec_2006} (2006)\\ \cline{5-7}
    81 & &  &  &  & (b) $T_e=260\,$eV & Chihara     & Gregori \textit{et al.}~\cite{Gregori_JQuantSpec_2006} (2006) \\ \cline{5-7}
    82 & &  &  &  & (c) $T_e=220\,$eV & Chihara     & Gregori \textit{et al.}~\cite{Gregori_JQuantSpec_2006} (2006) \\ \cline{5-7}
    83 & &  &  &  & (d) $T_e=200\,$eV & Chihara     & Gregori \textit{et al.}~\cite{Gregori_JQuantSpec_2006} (2006) \\ \hline
    84 & LiH (powder) shocked & Janus (LLNL) & $2.79\,$keV   & [Not Quoted]   &   $T\sim10-20\,$eV  & Detailed Balance     & Neumayer \textit{et al.}~\cite{Neumayer_RSI_2006} (2006)\\ \hline
    85 & C (foam) x-ray heated & OMEGA & $4.75\,$keV   & $130\pm5^\circ$    &  $T=52\,$eV ($\pm20\%$), $\rho=0.72\,$g/cc, $Z=4.25$ ($\pm20\%$)  & Chihara     & Gregori \textit{et al.}~\cite{Gregori_POP_2004} (2004)\\ \hline
    86 & C (foam) x-ray heated & OMEGA & $4.75\,$keV   & $130\pm5^\circ$    &  $T<5\,$eV , $\rho=0.72\,$g/cc, $Z=0.26$ ($\pm20\%$)  & Chihara     & Gregori \textit{et al.}~\cite{Gregori_POP_2004} (2004) \\ \hline
    87 & Be (solid) x-ray heated & OMEGA & $4.75\,$keV   & $125^\circ$    & $T\approx53\,$eV, $\rho=1.85\,$g/cc  & RPA     & Glenzer \textit{et al.}~\cite{Glenzer_PRL_2003,Glenzer_POP_2003} (2003) \\ \hline
    88 & LiH shocked & VULCAN & $4.75\,$keV   & $160^\circ$    & $T_e\sim0-5\,$eV, $n_e=1.8\pm0.6\times10^{23}$/cc, $Z=3.2\pm0.1$  & Chihara     & Gregori \textit{et al.}~\cite{Gregori_AIP_2002} (2002) \\ \hline
\end{longtblr}
\endgroup
\twocolumngrid
\end{widetext}

\begin{widetext}
\onecolumngrid
\begingroup
\small
\begin{longtblr}[
  caption = {Overview of XRTS experiments at pulsed power facilities ordered by year of publication.\label{table:pulsed}},
  label = {table:pulsed}
]{colspec = {Q[0.1,l] Q[1.5,l] Q[1.5,l] Q[1.2,l] Q[1.2,l] Q[3.0,l] Q[3.2,l]  Q[3.0,l]},
  rowhead = 1,
  hlines, columns={font=\footnotesize} } 
    & Material & Facility & $E_0$ & $\theta$ & conditions & method & reference \\
    \hline
    1 & Graphite (foil) pulsed power & Nevada TW & $4.75\,$keV   & $100\pm17^\circ$    & $T=2\pm1.9\,$eV, $\rho=0.6\pm0.5\,$g/cc  & Chihara     & Valenzuela \textit{et al.}~\cite{Valenzuela_SciRep_2018} (2018)\\
    2 & CH$_2$ (foam) shocked & Sandia Z-machine & $6.2\,$keV   & $\sim90^\circ$, $30\pm8^\circ$   & $T=4.3\,$eV, $\rho=0.52\,$g/cc,  $p=0.75\,$Mbar  & MHD     & Ao \textit{et al.}~\cite{AO201626} (2016) \\
    3 & CH$_2$ (foam) shocked & Sandia Z-machine & $6.2\,$keV   & $90^\circ$    &  $\rho=4\rho_0=0.4\,$g/cc  & ---     & Harding \textit{et al.}~\cite{Harding_RSI_2015} (2015) \\
    4 &  plastic (foam)  & Sandia Z-machine & $6.2\,$keV   & $90^\circ$   & ambient, $\rho=0.24\,$g/cc  & Chihara     & Golovkin \textit{et al.}~\cite{Golovkin_HEDP_2013} (2013) \\
\end{longtblr}
\endgroup
\twocolumngrid
\end{widetext}

\begin{widetext}
\onecolumngrid
\begingroup
\small
\begin{longtblr}[
  caption = {Overview of XRTS experiments at XFEL facilities (excluding DCA set-ups) ordered by year of publication.},
  label = {table:XFEL}
]{colspec = {X[0.1,l] X[0.9,l] X[0.5,l] X[1.2,l] X[1.0,l] X[1.4,l] X[1.4,l]  X[1.2,l]},
  rowhead = 1,
  columns={font=\footnotesize} } 
    & Material & Facility & $E_0$ & $\theta$ & conditions & method & reference \\
    \hline
    \hline
    1 & Cu (foil) ~~~ self heating & European XFEL & $8.75-9.9\,$keV SASE  &  $35^\circ, 170^\circ$   & $10\lesssim T_e \lesssim 150\,$eV  & ITCF   & \v{S}m{\'i}d \textit{et al.}~\cite{Smid_SciRep_2026} (2026)  \\ \hline    
    2 & Al (foil) shocked & European XFEL & $8.307\,$keV seeded  &  $10^\circ, 14^\circ, 18^\circ, 32^\circ$   & $T\approx0.6\,$eV, $3.75 \lesssim \rho \lesssim 4.5\,$g/cc  & DFT-MD/hydro LR-TDDFT   & Bespalov \textit{et al.}~\cite{bespalov2026experimentalevidencebreakdownuniformelectrongas} (2026)  \\ \hline
    3 & \SetCell[r=2]{h} Diamond  short pulse   & LCLS & $8.16\,$keV  SASE & $17^\circ$, $170^\circ$    &  $T=61\substack{+27 \\ -19}$\,eV (2PP foam, CHO data not diagnosed)  &   Chihara  & Martin \textit{et al.}~\cite{Martin_POP_2025} (2025)\\  \cline[dotted]{4-7}
    4 & &   &  & $170^\circ$    &  $T=85\pm13$\,eV  &   ITCF  & B\"ohme \textit{et al.}~\cite{bohme2026correlationfunctionmetrologywarm} (2026)\\ \hline
    5 & TMPTA (foam) shock& LCLS & $8.2\,$keV seeded  & $166^\circ$    &   $T\lesssim0.6\,$eV  & ---   & Heaton \textit{et al.}~\cite{Heaton_PRR_2025}  (2025) \\ \hline
    6 & Diamond & EurXFEL & $6\,$keV SASE+mono  & $18^\circ$    &   ambient  &  TDDFT   & Ranjan \textit{et al.}~\cite{Ranjan_POP_2023}  (2023) \\ \hline
    7 & Graphite ~~ self heating & LCLS & $5.9\,$keV SASE  & $29^\circ$    & $T=14\,$eV, $\rho=\rho_0=2.21\,$g/cc  & ITCF   & Bellenbaum \textit{et al.} (2025)~\cite{Bellenbaum_APL_2025} \\ \hline
    8 & Diamond (foil) & European XFEL & $6.0\,$keV SASE, SASE+mono  & $155^\circ$    & ambient  & ---   & Voigt \textit{et al.} (2021)~\cite{Voigt_POP_2021}; see also the f-sum rule evaluation by Dornheim \textit{et al.}~\cite{Dornheim_SciRep_2024} \\ \hline
    9 & Polystyrene shocked & LCLS & $8.18\,$keV SASE  & $17^\circ$, $124^\circ$    & $T\approx0.43\,$eV, $P\approx150\,$GPa  & Chihara / TDDFT / DFT-MD   & Frydrych \textit{et al.} (2020)~\cite{Frydrych_NatComm_2020} \\ \hline
    10 & \SetCell[r=3]{h} Graphite ~~ self heating  & LCLS & $5.9\,$keV SASE  & $160^\circ$    & $T=21.7$eV, $\rho=\rho_0=2.21\,$g/cc  & Chihara   & Kraus \textit{et al.}  (2019)~\cite{kraus_xrts}  \\ \cline[dotted]{5-7}
    11 & &   &   &   & $T=18\pm2\,$eV   & ITCF   & Dornheim \textit{et al.}~\cite{Dornheim_T_2022,Dornheim_T2_2022} (2022/23), Bellenbaum \emph{et al.}~\cite{Bellenbaum_APL_2025}~(2025) \\ \cline[dotted]{5-7}
    12 & &   &   &   &  $T\approx16.6\,$eV    & Chihara (free-bound)     &  B\"ohme \emph{et al.}~\cite{bohme2023evidencefreeboundtransitionswarm}  (2023)\\ \hline
    13 & Al (foil) shocked & LCLS & $7.2\,$keV seeded  & $10\pm4^\circ$, $36\pm4^\circ$    & $0.25\leq T\leq0.37\,$eV, $\rho\approx1.3\rho_0$  & plasmon dispersion   & Preston \textit{et al.}~\cite{Preston_APL_2019} (2019); see also TDDFT calculations by Ramakrishna \textit{et al.}~\cite{Ramakrishna_PRB_2021}\\ \hline
    14 & Graphite (foil) shocked & LCLS & $4.5/6.0\,$keV SASE  & $129^\circ$   & 6 data sets, $\rho=3.9-4.23$ and $T=0.45-0.8\,$eV  & Chihara / DFT-MD   & Helfrich \textit{et al.}~\cite{Helfrich_HEDP_2019} (2019)\\ \hline
    15 & Polystyrene shocked & LCLS & $8.2\,$keV SASE  & $17^\circ$, $123^\circ$    & ambient and shocked  & ---   & Kraus \textit{et al.}~\cite{Kraus_POP_2018} (2018)\\ \hline
    16 & Al (foil) ~~ self heating  & LCLS & $7.98\,$keV  seeded & $\theta=18^\circ$    & $T_e=0.3\,$eV, $\rho=\rho_0$  & DFT-MD Kubo-Greenwood & Witte \textit{et al.}~\cite{Witte_PRL_2017} (2017); see also TDDFT calculations by Ramakrishna \textit{et al.}~\cite{Ramakrishna_PRB_2021} \\ \hline
    17 & Diamond (foil) shocked & LCLS & $8\,$keV seeded  & $5^\circ < \theta < 30^\circ$, $60^\circ$    & $T=4\pm1\,$eV, $n_e=2\times10^{23}\,$/cc  & Chihara    & Lu \textit{et al.}~\cite{Lu_Spec_2017} (2017)  \\ \hline   
    18 & Diamond (foil) shocked & LCLS & $8\,$keV seeded  & $5^\circ < \theta < 30^\circ$, $60^\circ$    & $T\approx0.2\pm0.1\,$eV, $4\,$g/cc$<\rho<6.5\,$g/cc  & Chihara / DFT-MD   & Gamboa \textit{et al.}~\cite{Gamboa_POP_2015} (2015)  \\ \hline
    19 & Graphite & LCLS & $5.9\,$keV  SASE & $161^\circ$    & ambient  & ---   & Kraus \textit{et al.}~\cite{Kraus_POP_2015} (2015) \\ \hline
    20 & Al (foil) shocked & LCLS & $8\,$keV  seeded & $13^\circ$, $130^\circ$    & $T=1.75\pm0.5\,$eV, $\rho=6.3\,$g/cc  & Bohm-Gross / DFT-MD   & Fletcher \textit{et al.}~\cite{Fletcher2015} (2015) \\ \hline
    21 & \SetCell[r=3]{h} Al (foil) ~~ self heating & LCLS & $8\,$keV seeded  & $5^\circ < \theta < 30^\circ$, $60^\circ$    & $T=6\pm0.5\,$eV, $\rho=\rho_0$  & detailed balance & Sperling \textit{et al.}~\cite{Sperling_PRL_2015} (2015) \\ \cline[dotted]{5-7}
    22 &  &    &   & $T<2\,$eV, $\rho=\rho_0$  & LR-TDDFT   & Mo \textit{et al.}~\cite{Mo_PRL_2018} (2018) \\ \cline[dotted]{5-7}
    23 &  &    &   & $T=6.5\pm0.5\,$eV  &  ITCF  & Dornheim \textit{et al.}~\cite{Dornheim_T_2022}  (2022) \\ \hline
    24 & Al (foil) ~~ self heating & LCLS & $8\,$keV seeded  &  $13^\circ$, $130^\circ$   &  $\rho=2.7\,$g/cc, $Z=3$, low $T$  & plasmon dispersion   & Fletcher \textit{et al.}~\cite{Fletcher_RSI_2014} (2014)  \\ \hline
    25 & H (jet) ~~~ x-ray heating &  FLASH & $0.092\,$keV SASE    & $90^\circ$    & $T_e=14\pm3.5\,$eV, $n_e=2.6\pm0.2\times10^{20}$/cc  & Bohm-Gross ($n_e$), detailed balance ($T_e$) & Toleikis \textit{et al.}~\cite{Toleikis_2010} (2010) \\ \hline
    26 & H (jet) ~~~ self heating &  FLASH  & $0.092\,$keV SASE   & $90^\circ$    & $T_e=1-2\,$eV, $T_i\sim0.3-0.4\,$eV, $\rho=0.08\,$g/cc  & DFT-MD & Zastrau \textit{et al.}~\cite{Zastrau} (2010) \\ \hline
    27 & H (jet) ~~~ self heating &  FLASH  & $0.0918\,$keV SASE   & $90^\circ$    & $T=13\,$eV, $n_e=2.8\times10^{20}\,$/cc  & Chihara (non-equilibrium) & F\"austlin \textit{et al.}~\cite{Faustlin_PRL_2010} (2010); see also non-equilibrium analysis by Chapman and Gericke~\cite{Chapman_PRL_2011} \\ \hline
\end{longtblr}
\endgroup
\twocolumngrid
\end{widetext}

\begin{widetext}
\onecolumngrid
\begingroup
\small
\begin{longtblr}[
  caption = {Overview of XRTS experiments at XFEL facilities using DCA set-ups ordered by year of publication.},
  label = {table:DCA}
]{colspec = {X [0.1,l] X[0.9,l] X[0.75,l] X[1.5,l] X[1.2,l] X[1.4,l] X[1.2,l] X[1.0,l] X[1.0,l]},
  rowhead = 1,
  hlines, columns={font=\footnotesize} } 
    & Material & Facility & $E_0$ & $\theta$ & conditions & method & effect & reference \\
    \hline
    1 & Al (foil) & EurXFEL & $7.703\,$keV seeded/mono/DCA  & $25.6\pm1.4^\circ$  & ambient  & LR-TDDFT &  crystal orientation  $q$-averaging   & Gawne \textit{et al.}~\cite{gawne2025orientationaleffectslowpair} (2026)\\
    2 & Gold (foil) short-pulse & LCLS & $7.4919\,$keV  seeded/mono/DCA & $167^\circ-173^\circ$  &  $T=1.6\,$eV & Doppler broadening &  melting  entropy catastrophe  & White \textit{et al.}~\cite{White_Nature_2025} (2025) \\
    3 & Si (single crystal wafer) & EurXFEL & $7.703\,$keV  seeded/mono/DCA & $3.6^\circ, 8^\circ, 13.6^\circ$ $18.6^\circ, 25.6^\circ$ & ambient  & LR-TDDFT Bohm-Gross &  plasmon dispersion $q$-averaging, orientation   & Gawne \textit{et al.}~\cite{Gawne_ElectronicStructure_2025} (2025) \\
    4 & Al (foil) & EurXFEL & $7.703\,$keV seeded/mono/DCA   & $3.6^\circ, 8^\circ, 13.6^\circ$  $18.6^\circ, 25.6^\circ$ & ambient  & LR-TDDFT Bohm-Gross &  plasmon dispersion  $q$-averaging   & Gawne \textit{et al.}~\cite{Gawne_PRB_2024} (2024) \\
    5 & Methane (jet) laser heated & LCLS & $7.492\,$keV seeded/mono/DCA   & $47^\circ$ (no DCA)  $10^\circ, 20^\circ, 30^\circ$  & $T=0.3\pm0.1\,$eV   $\rho=0.8\pm0.1\,$g/cc   & Chihara &  sound speed   & White \textit{et al.}~\cite{White_PRR_2024} (2024) \\
    7 & Diamond  resistive heater & EurXFEL & $7.492\,$keV mono/DCA  & $8\pm1^\circ$, $12.2\pm1^\circ$  & ambient,  $T=500\,$K  & detailed balance &  phonon modes   & Descamps \textit{et al.}~\cite{Descamps_SciReports_2020} (2020)\\
\end{longtblr}
\endgroup
\twocolumngrid
\end{widetext}

\begin{widetext}
\onecolumngrid
\begingroup
\small
\begin{longtblr}[
  caption = {Overview of XRTS experiments at synchrotron and storage ring facilities ordered by year of publication. },
  label = {table:synchrotron}
]{colspec = {X [0.1,l] X[0.9,l] X[0.75,l] X[1.5,l] X[1.2,l] X[1.4,l] X[1.2,l] X[1.0,l] X[1.0,l]},
  rowhead = 1,
  columns={font=\footnotesize} } 
    & Material & Facility & $E_0$ & $\theta$ & conditions & method & effect & reference \\ 
    \hline
    \hline
    1 & Be & Advanced Photon Source (APS) & $9.89\,$keV, mono/spherical Si analyser & 54$^\circ-171^\circ$  & ambient & impulse approximation &  -- & Mattern \textit{et al.}~\cite{Mattern_PRB_2012} (2012); see also average-atoms calculations by Souza \textit{et al.}~\cite{Souza_PRE_2014}\\ \hline

    2 & H (diamond anvil cell) & Diamond Light Source, UK & $19.952\,$keV   & $32.2\pm1.1^\circ$  & ambient $T$\newline (a) $P=2.8\,$GPa \newline (b) $P=6.4\,$GPa  & reduction analysis (ratio to isolate H from C [DAC] scattering) &  dynamic LFC measurement   & Falk \textit{et al.}~\cite{Falk_JPhysConf_2010} (2010)\\ \hline

    3 & Si (single crystal wafer) & DORIS, HASYLAB, Germany & 7.99\,keV, mono/spherical Si analyser & 10.7$^\circ$--68.5$^\circ$  & ambient & RPA, local-pseudopotential &  Plasmon-Fano resonance & Sturm \textit{et al.}~\cite{Sturm_IXSS_1992} (2010)\\ \hline

     4 & Si (single crystal wafer) & ID16, ESRF, France & 7.9-8.0\,keV, mono/spherical Si analyser & 7.3$^\circ$--68.5$^\circ$  & ambient & LR-TDDFT &  Orientation, lifetime effects  & Weissker \textit{et al.}~\cite{Weissker_Lifetime_TDDFT} (2010)\\ \hline

     5 & Al (foil) & HARWI, HASYLAB, Germany & 13.7\,keV, mono/spherical Si analyser & 10.3$^\circ$ and 41.0$^\circ$  & $T=300$--1100\,K & RPA, local-pseudopotential &  Orientation & Sternemann \textit{et al.}~\cite{Sternemann_Heated_Al} (1998)\\ \hline

    6 & \SetCell[r=2]{h}Si (single crystal wafer) & HARWI, HASYLAB, Germany & $13.7\,$keV, mono/spherical Si analyser & 5.8$^\circ$--32.6$^\circ$  & ambient & RPA, LFC, on-shell corrections, local-pseudopotential &  Plasmon dispersion, orientation  & Sch\"ulke \textit{et al.}~\cite{Schuelke_Si} (1995) \\ \cline[dotted]{6-8}
    7 &  &  &  & &  & LR-TDDFT &  Orientation, lifetime effects, band gap  & Weissker \textit{et al.}~\cite{Weissker_Lifetime_TDDFT} (2010) \\ \hline
\end{longtblr}
\endgroup
\twocolumngrid
\end{widetext}

\section{Summary and Outlook\label{sec:outlook}}
Matter at extreme conditions is becoming increasingly important for basic science as well as a multitude of applications in astrophysics, material science, and, of course, inertial confinement fusion. This research field is bound to push the boundaries with respect to both experimentally creating extreme states of matter and diagnosing these samples with high accuracy and precision. In parallel, advances in the theoretical description and, especially, ab initio simulations propel the field forward and allow for new ways to interpret experimental data.

Future developments are foreshadowed by impressive technical advances in the last decade. 
On the diagnostic side, monochromators and new analyzer enable a new class of measurements with ultrahigh spectral resolution \cite{McBride_RSI_2018, Gawne_PRB_2024} that allows to resolve the collective ion motion. Similarly important is the emerging capability of pump-probe experiments with ultrahigh repetition rates \cite{Gorman_POP_2024} as such experiments will strongly improve the data quality. Furthermore, new statistical data analysis methods based on Bayesian inference or Markov-chain Monte Carlo \cite{Kasim_POP_2019, Poole_PPCF_2025} are replacing direct fits via minimizing a $\chi^2$-metric thereby also providing information on uncertainties. New and improved facilities will allow new setups and target types. A first example is the development of two-color schemes at XFEL facilities, which allow for x-ray pump and x-ray probe experiments with a very high temporal resolution \cite{SPERLING2011145}. Another example are experiments on matter heated by energetic ion beams, e.g., at the FAIR facility being built at GSI Darmstadt \cite{Fortov_FAIR_2012, Durante_2019, FAIR_PoP_2020}, that will employ larger targets with less gradients.

These experimental developments are complemented by important theoretical innovations, such as improved Chihara models \cite{bohme2023evidencefreeboundtransitionswarm}, model-free analysis methods \cite{Dornheim_T_2022, Dornheim_T2_2022, Dornheim_SciRep_2024} as well as new ab initio simulations based on both density functional theory \cite{Baczewski_PRL_2016, White_ElectronicStructure_2025, Moldabekov_MRE_2025,moldabekov2025enhancingefficiencytimedependentdensity} and path integral Monte Carlo \cite{Dornheim_NatComm_2025, schwalbe2025staticlineardensityresponse} schemes. New approaches include, e.g., attempts to construct the structure factor from path integral Monte Carlo simulations via analytic continuation \cite{Chuna_JPA_2025, chuna2025noiselesslimitimprovedpriorlimit, BENEDIXROBLES2026109904}. As ab initio simulations are computationally very expensive, representations of such data via machine-learning looks very attractive~\cite{dornheim_ML}. It would also allow for an efficient mixing of zones for samples with large gradients. These results are currently implemented into openly available codes \cite{xdave_code_paper,xdave} that are compatible with ray-tracing procedures and, thus, allow for an improved representation of the experimental details \cite{Gawne_CompPhysComm_2026}.

We are convinced that XRTS as one of the premier and well-established diagnostic methods will play an important part in future investigations of dense matter at high temperatures, be it to determine basic thermodynamic parameters or atomic-scale details of the electron and ion structure. Armed with this knowledge, questions about the behavior of other integral quantities, like the equation of state, stopping power, the mode structure or relaxation rates, can be addressed more rigorously. The rich information content of x-ray scattering spectra will also provide avenues for investigating high-energy-density matter beyond the current approaches and, thus, provide a much more detailed picture of this important parameter region.

\section*{Acknowledgements}

TD and DOG wish to thank the Lawrence Livermore National Laboratory (LLNL) for their hospitality during the final stages of this work.

This work has received funding from the European Union's Just Transition Fund (JTF) within the project \emph{R\"ontgenlaser-Optimierung der Laserfusion} (ROLF), contract number 5086999001, co-financed by the Saxon state government out of the State budget approved by the Saxon State Parliament. This work has received funding from the European Research Council (ERC) under the European Union’s Horizon 2022 research and innovation programme (Grant agreement No. 101076233, "PREXTREME"). 
Views and opinions expressed are however those of the authors only and do not necessarily reflect those of the European Union or the European Research Council Executive Agency. Neither the European Union nor the granting authority can be held responsible for them.
Tobias Dornheim gratefully acknowledges funding from the Deutsche Forschungsgemeinschaft (DFG) via project DO 2670/1-1.


\section*{Author Declarations}
\subsection*{Conflict of interest}

The authors have no conflicts to disclose.

\bibliography{bibliography}

@article{FAIR_PoP_2020,
    author = {Schoenberg, K. and Bagnoud, V. and Blazevic, A. and Fortov, V. E. and Gericke, D. O. and Golubev, A. and Hoffmann, D. H. H. and Kraus, D. and Lomonosov, I. V. and Mintsev, V. and Neff, S. and Neumayer, P. and Piriz, A. R. and Redmer, R. and Rosmej, O. and Roth, M. and Schenkel, T. and Sharkov, B. and Tahir, N. A. and Varentsov, D. and Zhao, Y.},
    title = {High-energy-density-science capabilities at the Facility for Antiproton and Ion Research},
    journal = {Physics of Plasmas},
    volume = {27},
    number = {4},
    pages = {043103},
    year = {2020},
    month = {04},
    issn = {1070-664X},
    doi = {10.1063/1.5134846},
    url = {https://doi.org/10.1063/1.5134846}
}

@article{Hall_PoP_2024,
    author = {Hall, G. N. and Weber, C. R. and Smalyuk, V. A. and Landen, O. L. and Trosseille, C. and Pak, A. and Hartouni, E. and Marley, E. and Ebert, T. and Bradley, D. K. and Hsing, W. and Tommasini, R. and Izumi, N. and Pape, S. Le and Divol, L. and Krauland, C. M. and Thompson, N. and Casco, E. R. and Ayers, M. J. and Nagel, S. R. and Carpenter, A. C. and Hurd, E. R. and Dayton, M. S. and Engelhorn, K. and Holder, J. P.},
    title = {Measurement of mix at the fuel–ablator interface in indirectly driven capsule implosions on the {N}ational {I}gnition {F}acility},
    journal = {Physics of Plasmas},
    volume = {31},
    number = {2},
    pages = {022702},
    year = {2024},
    month = {02},
    issn = {1070-664X},
    doi = {10.1063/5.0171617},
    url = {https://doi.org/10.1063/5.0171617}
}

@article{Maybey_NatComm_2017,
  title = {A strong diffusive ion mode in dense ionized matter predicted by Langevin dynamics},
  author = {Mabey, P. and Richardson, S. and White, T. G. and Fletcher, L. B. and Glenzer, S. H. and Hartley, N. J. and Vorberger, J. and Gericke, D. O. and Gregori, G.},
  journal = {Nature Communications},
  volume = {8},
  issue = {1},
  pages = {14125},
  year = {2017},
  doi = {10.1038/ncomms14125},
  url = {https://doi.org/10.1038/ncomms14125}
}

@article{Poole_PoP_2022,
    author = {Poole, H. and Cao, D. and Epstein, R. and Golovkin, I. and Walton, T. and Hu, S. X. and Kasim, M. and Vinko, S. M. and Rygg, J. R. and Goncharov, V. N. and Gregori, G. and Regan, S. P.},
    title = {A case study of using x-ray {T}homson scattering to diagnose the in-flight plasma conditions of DT cryogenic implosions},
    journal = {Physics of Plasmas},
    volume = {29},
    number = {7},
    pages = {072703},
    year = {2022},
    month = {07},
    issn = {1070-664X},
    doi = {10.1063/5.0072790},
    url = {https://doi.org/10.1063/5.0072790}
    }

@article{Kraus_Nature_2025,
author = {Kraus, D. and Rips, J. and Schörner, M. and Stevenson, M. G. and Vorberger, J. and Ranjan, D. and Lütgert, J. and Heuser, B. and Eggert, J. H. and Liermann, H. -P. and Oleynik, I. I. and Pandolfi, S. and Redmer, R. and Sollier, A. and Strohm, C. and Volz, T. J. and Albertazzi, B. and Ali, S. J. and Antonelli, L. and Bähtz, C. and Ball, O. B. and Banerjee, S. and Belonoshko, A. B. and Bolme, C. A. and Bouffetier, V. and Briggs, R. and Buakor, K. and Butcher, T. and Cerantola, V. and Chantel, J. and Coleman, A. L. and Collier, J. and Collins, G. W. and Comley, A. J. and Cowan, T. E. and Cristoforetti, G. and Cynn, H. and Descamps, A. and Di Cicco, A. and Di Dio Cafiso, S. and Dorchies, F. and Duff, M. J. and Dwivedi, A. and Edwards, C. and Errandonea, D. and Galitskiy, S. and Galtier, E. and Ginestet, H. and Gizzi, L. and Gleason, A. and Göde, S. and Gonzalez, J. M. and Gorman, M. G. and Harmand, M. and Hartley, N. J. and Heighway, P. G. and Hernandez-Gomez, C. and Higginbotham, A. and Höppner, H. and Husband, R. J. and Hutchinson, T. M. and Hwang, H. and Keen, D. A. and Kim, J. and Koester, P. and Konôpková, Z. and Krygier, A. and Labate, L. and Laso Garcia, A. and Lazicki, A. E. and Lee, Y. and Mason, P. and Masruri, M. and Massani, B. and McBride, E. E. and McHardy, J. D. and McGonegle, D. and McGuire, C. and McWilliams, R. S. and Merkel, S. and Morard, G. and Nagler, B. and Nakatsutsumi, M. and Nguyen-Cong, K. and Norton, A. -M. and Ozaki, N. and Otzen, C. and Peake, D. J. and Pelka, A. and Pereira, K. A. and Phillips, J. P. and Prescher, C. and Preston, T. R. and Randolph, L. and Ravasio, A. and Santamaria-Perez, D. and Savage, D. J. and Schölmerich, M. and Schwinkendorf, J. -P. and Singh, S. and Smith, J. and Smith, R. F. and Spear, J. and Spindloe, C. and Suer, T. -A. and Tang, M. and Toncian, M. and Toncian, T. and Tracy, S. J. and Trapananti, A. and Vennari, C. E. and Vinci, T. and Tyldesley, M. and Vogel, S. C. and Walsh, J. P. S. and Wark, J. S. and Willman, J. T. and Wollenweber, L. and Zastrau, U. and Brambrink, E. and Appel, K. and McMahon, M. I.},
title = {The structure of liquid carbon elucidated by in situ X-ray diffraction},
journal = {Nature},
volume = {642},
number = {8067},
pages = {351--355},
year = {2025},
doi = {10.1038/s41586-025-09035-6},
URL = {https://doi.org/10.1038/s41586-025-09035-6},
}

@article{VERBENI_2005,
title = {Advances in crystal analyzers for inelastic X-ray scattering},
journal = {Journal of Physics and Chemistry of Solids},
volume = {66},
number = {12},
pages = {2299-2305},
year = {2005},
note = {5th International Conference on Inelastic X-ray Scattering (IXS 2004)},
issn = {0022-3697},
doi = {https://doi.org/10.1016/j.jpcs.2005.09.079},
url = {https://www.sciencedirect.com/science/article/pii/S0022369705003598},
author = {R. Verbeni and M. Kocsis and S. Huotari and M. Krisch and G. Monaco and F. Sette and G. Vanko}
}

@article{Descamps_2022,
author = "Descamps, A. and Ofori-Okai, B. K. and Baldwin, J. K. and Chen, Z. and Fletcher, L. B. and Glenzer, S. H. and Hartley, N. J. and Hasting, J. B. and Khaghani, D. and Mo, M. and Nagler, B. and Recoules, V. and Redmer, R. and Sch{\"{o}}rner, M. and Sun, P. and Wang, Y. Q. and White, T. G. and McBride, E. E.",
title = "{Towards performing high-resolution inelastic X-ray scattering measurements at hard X-ray free-electron lasers coupled with energetic laser drivers}",
journal = "Journal of Synchrotron Radiation",
year = "2022",
volume = "29",
number = "4",
pages = "931--938",
month = "Jul",
doi = {10.1107/S1600577522004453},
url = {https://doi.org/10.1107/S1600577522004453},
}

@article{Vorberger_PRL_2012,
  title = {Dynamic Ion Structure Factor of Warm Dense Matter},
  author = {Vorberger, J. and Donko, Z. and Tkachenko, I. M. and Gericke, D. O.},
  journal = {Phys. Rev. Lett.},
  volume = {109},
  issue = {22},
  pages = {225001},
  numpages = {5},
  year = {2012},
  month = {Nov},
  publisher = {American Physical Society},
  doi = {10.1103/PhysRevLett.109.225001},
  url = {https://link.aps.org/doi/10.1103/PhysRevLett.109.225001}
}

@article{Gregori_PoP_2009,
    author = {Gregori, G. and Gericke, D. O.},
    title = {Low frequency structural dynamics of warm dense matter},
    journal = {Physics of Plasmas},
    volume = {16},
    number = {5},
    pages = {056306},
    year = {2009},
    month = {04},
    issn = {1070-664X},
    doi = {10.1063/1.3100203},
    url = {https://doi.org/10.1063/1.3100203}
}

@article{Lutgert_PoP_2022,
    author = {Lütgert, J. and Bethkenhagen, M. and Bachmann, B. and Divol, L. and Gericke, D. O. and Glenzer, S. H. and Hall, G. N. and Izumi, N. and Khan, S. F. and Landen, O. L. and MacLaren, S. A. and Masse, L. and Redmer, R. and Schörner, M. and Schölmerich, M. O. and Schumacher, S. and Shaffer, N. R. and Starrett, C. E. and Sterne, P. A. and Trosseille, C. and Döppner, T. and Kraus, D.},
    title = {Platform for probing radiation transport properties of hydrogen at
                    conditions found in the deep interiors of red dwarfs},
    journal = {Physics of Plasmas},
    volume = {29},
    number = {8},
    pages = {083301},
    year = {2022},
    month = {08},
    issn = {1070-664X},
    doi = {10.1063/5.0094579},
    url = {https://doi.org/10.1063/5.0094579}
}

@article{Bishel_RSI_2018,
    author = {Bishel, D. T. and Bachmann, B. and Yi, A. and Kraus, D. and Divol, L. and Bethkenhagen, M. and Falcone, R. W. and Fletcher, L. B. and Glenzer, S. H. and Landen, O. L. and MacDonald, M. J. and Masters, N. and Neumayer, P. and Redmer, R. and Saunders, A. M. and Witte, B. B. L. and Döppner, T.},
    title = {Using time-resolved penumbral imaging to measure low hot spot x-ray emission signals from capsule implosions at the {N}ational {I}gnition {F}acility},
    journal = {Review of Scientific Instruments},
    volume = {89},
    number = {10},
    pages = {10G111},
    year = {2018},
    month = {09},
    issn = {0034-6748},
    doi = {10.1063/1.5037073},
    url = {https://doi.org/10.1063/1.5037073}
    }

@article{Gericke_PRE_2010,
  title = {Screening of ionic cores in partially ionized plasmas within linear response},
  author = {Gericke, D. O. and Vorberger, J. and W\"unsch, K. and Gregori, G.},
  journal = {Phys. Rev. E},
  volume = {81},
  issue = {6},
  pages = {065401},
  numpages = {4},
  year = {2010},
  month = {Jun},
  publisher = {American Physical Society},
  doi = {10.1103/PhysRevE.81.065401},
  url = {https://link.aps.org/doi/10.1103/PhysRevE.81.065401}
}

@article{Wunsch_PRE_2009,
  title = {Ion structure in warm dense matter: Benchmarking solutions of hypernetted-chain equations by first-principle simulations},
  author = {W\"unsch, K. and Vorberger, J. and Gericke, D. O.},
  journal = {Phys. Rev. E},
  volume = {79},
  issue = {1},
  pages = {010201},
  numpages = {4},
  year = {2009},
  month = {Jan},
  publisher = {American Physical Society},
  doi = {10.1103/PhysRevE.79.010201},
  url = {https://link.aps.org/doi/10.1103/PhysRevE.79.010201}
}

@article{Ma_PoP_2014,
    author = {Ma, T. and Fletcher, L. and Pak, A. and Chapman, D. A. and Falcone, R. W. and Fortmann, C. and Galtier, E. and Gericke, D. O. and Gregori, G. and Hastings, J. and Landen, O. L. and Le Pape, S. and Lee, H. J. and Nagler, B. and Neumayer, P. and Turnbull, D. and Vorberger, J. and White, T. G. and Wünsch, K. and Zastrau, U. and Glenzer, S. H. and Döppner, T.},
    title = {Observations of strong ion-ion correlations in dense plasmas},
    journal = {Physics of Plasmas},
    volume = {21},
    number = {5},
    pages = {056302},
    year = {2014},
    month = {04},
    issn = {1070-664X},
    doi = {10.1063/1.4872161},
    url = {https://doi.org/10.1063/1.4872161}
    }

@article{Glenzer_1999,
  title = {{T}homson Scattering from High-$\mathit{Z}$ Laser-Produced Plasmas},
  author = {Glenzer, S. H. and Rozmus, W. and MacGowan, B. J. and Estabrook, K. G. and De Groot, J. D. and Zimmerman, G. B. and Baldis, H. A. and Harte, J. A. and Lee, R. W. and Williams, E. A. and Wilson, B. G.},
  journal = {Phys. Rev. Lett.},
  volume = {82},
  issue = {1},
  pages = {97--100},
  numpages = {0},
  year = {1999},
  month = {Jan},
  publisher = {American Physical Society},
  doi = {10.1103/PhysRevLett.82.97},
  url = {https://link.aps.org/doi/10.1103/PhysRevLett.82.97}
}

@article{Vorberger_2015,
  title = {Ab initio approach to model x-ray diffraction in warm dense matter},
  author = {Vorberger, J. and Gericke, D. O.},
  journal = {Phys. Rev. E},
  volume = {91},
  issue = {3},
  pages = {033112},
  numpages = {5},
  year = {2015},
  month = {Mar},
  publisher = {American Physical Society},
  doi = {10.1103/PhysRevE.91.033112},
  url = {https://link.aps.org/doi/10.1103/PhysRevE.91.033112}
}

@article{Cross_2016,
  title = {Theory of density fluctuations in strongly radiative plasmas},
  author = {Cross, J. E. and Mabey, P. and Gericke, D. O. and Gregori, G.},
  journal = {Phys. Rev. E},
  volume = {93},
  issue = {3},
  pages = {033201},
  numpages = {7},
  year = {2016},
  month = {Mar},
  publisher = {American Physical Society},
  doi = {10.1103/PhysRevE.93.033201},
  url = {https://link.aps.org/doi/10.1103/PhysRevE.93.033201}
}

@article{Bornath_2019,
  title = {{T}homson scattering from dense inhomogeneous plasmas},
  author = {Beuermann, T.-N. and Redmer, R. and Bornath, Th.},
  journal = {Phys. Rev. E},
  volume = {99},
  issue = {5},
  pages = {053205},
  numpages = {9},
  year = {2019},
  month = {May},
  publisher = {American Physical Society},
  doi = {10.1103/PhysRevE.99.053205},
  url = {https://link.aps.org/doi/10.1103/PhysRevE.99.053205}
}

@article{LANDEN2024101102,
title = {Probing dense plasmas for {HEDS} and {ICF}},
journal = {High Energy Density Physics},
volume = {51},
pages = {101102},
year = {2024},
issn = {1574-1818},
doi = {https://doi.org/10.1016/j.hedp.2024.101102},
url = {https://www.sciencedirect.com/science/article/pii/S1574181824000272},
author = {O L Landen}
}

@article{Daligault_2009,
doi = {10.1088/0004-637X/703/1/994},
url = {https://doi.org/10.1088/0004-637X/703/1/994},
year = {2009},
month = {sep},
publisher = {The American Astronomical Society},
volume = {703},
number = {1},
pages = {994},
author = {Daligault, J. and Gupta, S.},
title = {Electron-ion scattering in dense multi-component plasmas: applications to the outer crust of an accreting neutron star},
journal = {The Astrophysical Journal}
}

@article{zylstra_Nature_2022,
  title = {Burning Plasma Achieved in Inertial Fusion},
  author = {Zylstra, A. B. and Hurricane, O. A. and Callahan, D. A. and Kritcher, A. L. and Ralph, J. E. and Robey, H. F. and Ross, J. S. and Young, C. V. and Baker, K. L. and Casey, D. T. and D{\"o}ppner, T. and Divol, L. and Hohenberger, M. and Le Pape, S. and Pak, A. and Patel, P. K. and Tommasini, R. and Ali, S. J. and Amendt, P. A. and Atherton, L. J. and Bachmann, B. and Bailey, D. and Benedetti, L. R. and Berzak Hopkins, L. and Betti, R. and Bhandarkar, S. D. and Biener, J. and Bionta, R. M. and Birge, N. W. and Bond, E. J. and Bradley, D. K. and Braun, T. and Briggs, T. M. and Bruhn, M. W. and Celliers, P. M. and Chang, B. and Chapman, T. and Chen, H. and Choate, C. and Christopherson, A. R. and Clark, D. S. and Crippen, J. W. and Dewald, E. L. and Dittrich, T. R. and Edwards, M. J. and Farmer, W. A. and Field, J. E. and Fittinghoff, D. and Frenje, J. and Gaffney, J. and Gatu Johnson, M. and Glenzer, S. H. and Grim, G. P. and Haan, S. and Hahn, K. D. and Hall, G. N. and Hammel, B. A. and Harte, J. and Hartouni, E. and Heebner, J. E. and Hernandez, V. J. and Herrmann, H. and Herrmann, M. C. and Hinkel, D. E. and Ho, D. D. and Holder, J. P. and Hsing, W. W. and Huang, H. and Humbird, K. D. and Izumi, N. and Jarrott, L. C. and Jeet, J. and Jones, O. and Kerbel, G. D. and Kerr, S. M. and Khan, S. F. and Kilkenny, J. and Kim, Y. and Geppert Kleinrath, H. and Geppert Kleinrath, V. and Kong, C. and Koning, J. M. and Kroll, J. J. and Kruse, M. K. G. and Kustowski, B. and Landen, O. L. and Langer, S. and Larson, D. and Lemos, N. C. and Lindl, J. D. and Ma, T. and MacDonald, M. J. and MacGowan, B. J. and Mackinnon, A. J. and MacLaren, S. A. and MacPhee, A. G. and Marinak, M. M. and Mariscal, D. A. and Marley, E. V. and Masse, L. and Meaney, K. and Meezan, N. B. and Michel, P. A. and Millot, M. and Milovich, J. L. and Moody, J. D. and Moore, A. S. and Morton, J. W. and Murphy, T. and Newman, K. and Di Nicola, J.-M. G. and Nikroo, A. and Nora, R. and Patel, M. V. and Pelz, L. J. and Peterson, J. L. and Ping, Y. and Pollock, B. B. and Ratledge, M. and Rice, N. G. and Rinderknecht, H. and Rosen, M. and Rubery, M. S. and Salmonson, J. D. and Sater, J. and Schiaffino, S. and Schlossberg, D. J. and Schneider, M. B. and Schroeder, C. R. and Scott, H. A. and Sepke, S. M. and Sequoia, K. and Sherlock, M. W. and Shin, S. and Smalyuk, V. A. and Spears, B. K. and Springer, P. T. and Stadermann, M. and Stoupin, S. and Strozzi, D. J. and Suter, L. J. and Thomas, C. A. and Town, R. P. J. and Tubman, E. R. and Trosseille, C. and Volegov, P. L. and Weber, C. R. and Widmann, K. and Wild, C. and Wilde, C. H. and Van Wonterghem, B. M. and Woods, D. T. and Woodworth, B. N. and Yamaguchi, M. and Yang, S. T. and Zimmerman, G. B.},
  year = 2022,
  month = jan,
  journal = {Nature},
  volume = {601},
  number = {7894},
  pages = {542--548},
  issn = {1476-4687},
  doi = {10.1038/s41586-021-04281-w}
}

@article{SAUMON20221,
title = {Current challenges in the physics of white dwarf stars},
journal = {Physics Reports},
volume = {988},
pages = {1-63},
year = {2022},
note = {Current Challenges in the Physics of White Dwarf Stars},
issn = {0370-1573},
doi = {https://doi.org/10.1016/j.physrep.2022.09.001},
url = {https://www.sciencedirect.com/science/article/pii/S0370157322003180},
author = {Didier Saumon and Simon Blouin and Pier-Emmanuel Tremblay}
}

@article{Kritcher_PRL_2011,
  title = {In-Flight Measurements of Capsule Shell Adiabats in Laser-Driven Implosions},
  author = {Kritcher, A. L. and D\"oppner, T. and Fortmann, C. and Ma, T. and Landen, O. L. and Wallace, R. and Glenzer, S. H.},
  journal = {Phys. Rev. Lett.},
  volume = {107},
  issue = {1},
  pages = {015002},
  numpages = {4},
  year = {2011},
  month = {Jul},
  publisher = {American Physical Society},
  doi = {10.1103/PhysRevLett.107.015002},
  url = {https://link.aps.org/doi/10.1103/PhysRevLett.107.015002}
}

@article{Ravasio_PRL_2007,
  title = {Direct Observation of Strong Ion Coupling in Laser-Driven Shock-Compressed Targets},
  author = {Ravasio, A. and Gregori, G. and Benuzzi-Mounaix, A. and Daligault, J. and Delserieys, A. and Faenov, A. Ya. and Loupias, B. and Ozaki, N. and Rabec le Gloahec, M. and Pikuz, T. A. and Riley, D. and Koenig, M.},
  journal = {Phys. Rev. Lett.},
  volume = {99},
  issue = {13},
  pages = {135006},
  numpages = {4},
  year = {2007},
  month = {Sep},
  publisher = {American Physical Society},
  doi = {10.1103/PhysRevLett.99.135006},
  url = {https://link.aps.org/doi/10.1103/PhysRevLett.99.135006}
}

@Article{Dornheim_NatComm_2025,
author={Dornheim, Tobias
and D{\"o}ppner, Tilo
and Tolias, Panagiotis
and B{\"o}hme, Maximilian P.
and Fletcher, Luke B.
and Gawne, Thomas
and Graziani, Frank R.
and Kraus, Dominik
and MacDonald, Michael J.
and Moldabekov, Zhandos A.
and Schwalbe, Sebastian
and Gericke, Dirk O.
and Vorberger, Jan},
title={Unraveling electronic correlations in warm dense quantum plasmas},
journal={Nature Communications},
year={2025},
month={Jun},
day={02},
volume={16},
number={1},
pages={5103},
issn={2041-1723},
doi={10.1038/s41467-025-60278-3},
url={https://doi.org/10.1038/s41467-025-60278-3}
}

@Article{Dornheim_SciRep_2024,
  author  = {T. Dornheim and T. Döppner and A. D. Baczewski and P. Tolias and M. P. Böhme and {Zh. A.} Moldabekov and {Th.} Gawne and D. Ranjan and D. A. Chapman and M. J. MacDonald and {Th. R.} Preston and D. Kraus and J. Vorberger},
  journal = {Scientific Reports},
  title   = {{X-ray {T}homson scattering absolute intensity from the f-sum rule in the imaginary-time domain}},
  year    = {2024},
  pages   = {14377},
  volume  = {14},
  doi     = {https://doi.org/10.1038/s41598-024-64182-6},
}

@article{Dornheim_MRE_2023,
  author = {Dornheim, Tobias and Moldabekov, Zhandos and Tolias, Panagiotis and Böhme, Maximilian and Vorberger, Jan},  
  title = {Physical insights from imaginary-time density--density correlation functions},
  journal = {Matter and Radiation at Extremes},
  volume = {8},
  pages = {056601},
  year = {2023},
  publisher = {American Physical Society},
  doi = {10.1063/5.0149638},
  url = {https://doi.org/10.1063/5.0149638}
}

@article{Bellenbaum_APL_2025,
    author = {Bellenbaum, H. M. and Bachmann, B. and Kraus, D. and Gawne, Th. and Böhme, M. P. and Döppner, T. and Fletcher, L. B. and MacDonald, M. J. and Moldabekov, Zh. A. and Preston, T. R. and Vorberger, J. and Dornheim, T.},
    title = {Toward model-free temperature diagnostics of warm dense matter from multiple scattering angles},
    journal = {Applied Physics Letters},
    volume = {126},
    number = {4},
    pages = {044104},
    year = {2025},
    month = {01},
    issn = {0003-6951},
    doi = {10.1063/5.0248230},
    url = {https://doi.org/10.1063/5.0248230}
    }

@article{Vorberger_PLA_2024,
title = {Revealing non-equilibrium and relaxation in laser heated matter},
journal = {Physics Letters A},
volume = {499},
pages = {129362},
year = {2024},
issn = {0375-9601},
doi = {https://doi.org/10.1016/j.physleta.2024.129362},
url = {https://www.sciencedirect.com/science/article/pii/S0375960124000574},
author = {Jan Vorberger and Thomas R. Preston and Nikita Medvedev and Maximilian P. Böhme and Zhandos A. Moldabekov and Dominik Kraus and Tobias Dornheim}
}

@misc{gawne2025spectraldeconvolutiondeconvolutionextracting,
      title={Spectral Deconvolution without the Deconvolution: Extracting Temperature from X-ray {T}homson Scattering Spectra without the Source-and-Instrument Function}, 
      author={Thomas Gawne and Alina Kononov and Andrew Baczewski and Hannah Bellenbaum and Maximilian P Böhme and Zhandos Moldabekov and Thomas R Preston and Sebastian Schwalbe and Jan Vorberger and Tobias Dornheim},
      year={2025},
      eprint={2510.26747},
      archivePrefix={arXiv},
      primaryClass={physics.plasm-ph},
      url={https://arxiv.org/abs/2510.26747}, 
}

@misc{gawne2025orientationaleffectslowpair,
      title={Orientational Effects in the Low Pair Continuum of Aluminium}, 
      author={Thomas Gawne and Zhandos A Moldabekov and Oliver S Humphries and Motoaki Nakatsutsumi and Sebastian Schwalbe and Jan Vorberger and Ulf Zastrau and Tobias Dornheim and Thomas R Preston},
      year={2025},
      eprint={2508.02251},
      archivePrefix={arXiv},
      primaryClass={cond-mat.mtrl-sci},
      url={https://arxiv.org/abs/2508.02251}, 
}

@misc{acosta2026montecarloeventgenerationxray,
      title={Monte-Carlo Event Generation for X-Ray Thomson Scattering Analysis}, 
      author={Uwe Hernandez Acosta and Thomas Gawne and Jan Vorberger and Hannah Bellenbaum and Anton Reinhard and Simeon Ehrig and Klaus Steiniger and Michael Bussmann and Tobias Dornheim},
      year={2026},
      eprint={2604.05935},
      archivePrefix={arXiv},
      primaryClass={hep-ph},
      url={https://arxiv.org/abs/2604.05935}, 
}

@book{sheffield2010plasma,
  title={Plasma Scattering of Electromagnetic Radiation: Theory and Measurement Techniques},
  author={Sheffield, J. and Froula, D. and Glenzer, S.H. and Luhmann, N.C.},
  isbn={9780080952031},
  url={https://books.google.de/books?id=1NS5Fxam1lkC},
  year={2010},
  publisher={Elsevier Science}
}

@article{Gawne_JAP_2024,
    author = {Gawne, Thomas and Bellenbaum, Hannah and Fletcher, Luke B. and Appel, Karen and Baehtz, Carsten and Bouffetier, Victorien and Brambrink, Erik and Brown, Danielle and Cangi, Attila and Descamps, Adrien and Goede, Sebastian and Hartley, Nicholas J. and Herbert, Marie-Luise and Hesselbach, Philipp and Höppner, Hauke and Humphries, Oliver S. and Konôpková, Zuzana and Laso Garcia, Alejandro and Lindqvist, Björn and Lütgert, Julian and MacDonald, Michael J. and Makita, Mikako and Martin, Willow and Mishchenko, Mikhail and Moldabekov, Zhandos A. and Nakatsutsumi, Motoaki and Naedler, Jean-Paul and Neumayer, Paul and Pelka, Alexander and Qu, Chongbing and Randolph, Lisa and Rips, Johannes and Toncian, Toma and Vorberger, Jan and Wollenweber, Lennart and Zastrau, Ulf and Kraus, Dominik and Preston, Thomas R. and Dornheim, Tobias},
    title = {Effects of mosaic crystal instrument functions on x-ray {T}homson scattering diagnostics},
    journal = {Journal of Applied Physics},
    volume = {136},
    number = {10},
    pages = {105902},
    year = {2024},
    month = {09},
    issn = {0021-8979},
    doi = {10.1063/5.0222072},
    url = {https://doi.org/10.1063/5.0222072}
    }

@article{Gawne_PRB_2024,
  title = {Ultrahigh resolution x-ray {T}homson scattering measurements at the {E}uropean {X}-ray {F}ree {E}lectron {L}aser},
  author = {Gawne, Thomas and Moldabekov, Zhandos A. and Humphries, Oliver S. and Appel, Karen and Baehtz, Carsten and Bouffetier, Victorien and Brambrink, Erik and Cangi, Attila and G\"ode, Sebastian and Kon\^opkov\'a, Zuzana and Makita, Mikako and Mishchenko, Mikhail and Nakatsutsumi, Motoaki and Ramakrishna, Kushal and Randolph, Lisa and Schwalbe, Sebastian and Vorberger, Jan and Wollenweber, Lennart and Zastrau, Ulf and Dornheim, Tobias and Preston, Thomas R.},
  journal = {Phys. Rev. B},
  volume = {109},
  issue = {24},
  pages = {L241112},
  numpages = {7},
  year = {2024},
  month = {Jun},
  publisher = {American Physical Society},
  doi = {10.1103/PhysRevB.109.L241112},
  url = {https://link.aps.org/doi/10.1103/PhysRevB.109.L241112}
}

@misc{vorberger2025roadmapwarmdensematter,
      title={Roadmap for warm dense matter physics}, 
      author={Jan Vorberger and Frank Graziani and David Riley and Andrew D. Baczewski and Isabelle Baraffe and Mandy Bethkenhagen and Simon Blouin and Maximilian P. Böhme and Michael Bonitz and Michael Bussmann and Alexis Casner and Witold Cayzac and Peter Celliers and Gilles Chabrier and Nicolas Chamel and Dave Chapman and Mohan Chen and Jean Clérouin and Gilbert Collins and Federica Coppari and Tilo Döppner and Tobias Dornheim and Luke B. Fletcher and Dirk O. Gericke and Siegfried Glenzer and Alexander F. Goncharov and Gianluca Gregori and Sebastien Hamel and Stephanie B. Hansen and Nicholas J. Hartley and Suxing Hu and Omar A. Hurricane and Valentin V. Karasiev and Joshua J. Kas and Brendan Kettle and Thomas Kluge and Marcus D. Knudson and Alina Kononov and Zuzana Konôpkov á and Dominik Kraus and Andrea Kritcher and Sophia Malko and Gérard Massacrier and Burkhard Militzer and Zhandos A. Moldabekov and Michael S. Murillo and Bob Nagler and Nadine Nettelmann and Paul Neumayer and Benjamin K. Ofori-Okai and Ivan I. Oleynik and Martin Preising and Aurora Pribram-Jones and Tlekkabul Ramazanov and Alessandra Ravasio and Ronald Redmer and Baerbel Rethfeld and Alex P. L. Robinson and Gerd Röpke and François Soubiran and Charles E. Starrett and Gerd Steinle-Neumann and Phillip A. Sterne and Shigenori Tanaka and Aidan P. Thompson and Samuel B. Trickey and Tommaso Vinci and Sam M. Vinko and Lei Wang and Alexander J. White and Thomas G. White and Ulf Zastrau and Eva Zurek and Panagiotis Tolias},
      year={2025},
      eprint={2505.02494},
      archivePrefix={arXiv},
      primaryClass={physics.plasm-ph},
      url={https://arxiv.org/abs/2505.02494}, 
}

@misc{kononov2025realtimetimedependentdensityfunctional,
      title={Real-time time-dependent density functional theory for high-energy density physics}, 
      author={Alina Kononov and Minh Nguyen and Andrew D. Baczewski},
      year={2025},
      eprint={2511.14643},
      archivePrefix={arXiv},
      primaryClass={physics.plasm-ph},
      url={https://arxiv.org/abs/2511.14643}, 
}

@article{Chuna_JPA_2025,
  title={Dual formulation of the maximum entropy method applied to analytic continuation of quantum {M}onte {C}arlo data},
  author={Chuna, Thomas and Barnfield, Nicholas and Dornheim, Tobias and Friedlander, Michael P and Hoheisel, Tim},
  journal={Journal of Physics A: Mathematical and Theoretical},
  volume={58},
  number={33},
  pages={335203},
  year={2025},
  doi={10.1088/1751-8121/adf924}
}

@article{Jarrell_PhysRep_1996,
title = {Bayesian inference and the analytic continuation of imaginary-time quantum {M}onte {C}arlo data},
journal = {Physics Reports},
volume = {269},
number = {3},
pages = {133-195},
year = {1996},
issn = {0370-1573},
doi = {https://doi.org/10.1016/0370-1573(95)00074-7},
url = {https://www.sciencedirect.com/science/article/pii/0370157395000747},
author = {Mark Jarrell and J.E. Gubernatis}
}

@article{Bellenbaum_PRR_2025,
  title = {Estimating ionization states and continuum lowering from ab initio path integral {M}onte {C}arlo simulations for warm dense hydrogen},
  author = {Bellenbaum, Hannah M. and B\"ohme, Maximilian P. and Bonitz, Michael and D\"oppner, Tilo and Fletcher, Luke B. and Gawne, Thomas and Kraus, Dominik and Moldabekov, Zhandos A. and Schwalbe, Sebastian and Vorberger, Jan and Dornheim, Tobias},
  journal = {Phys. Rev. Res.},
  volume = {7},
  issue = {3},
  pages = {033016},
  numpages = {21},
  year = {2025},
  month = {Jul},
  publisher = {American Physical Society},
  doi = {10.1103/9d7r-1xbm},
  url = {https://link.aps.org/doi/10.1103/9d7r-1xbm}
}

@misc{baczewski2021predictionsboundboundtransitionsignatures,
      title={Predictions of bound-bound transition signatures in x-ray {T}homson scattering}, 
      author={Andrew D. Baczewski and Thomas Hentschel and Alina Kononov and Stephanie B. Hansen},
      year={2021},
      eprint={2109.09576},
      archivePrefix={arXiv},
      primaryClass={physics.plasm-ph},
      url={https://arxiv.org/abs/2109.09576}, 
}

@article{Hentschel_POP_2023,
    author = {Hentschel, Thomas W. and Kononov, Alina and Olmstead, Alexandra and Cangi, Attila and Baczewski, Andrew D. and Hansen, Stephanie B.},
    title = {Improving dynamic collision frequencies: Impacts on dynamic structure factors and stopping powers in warm dense matter},
    journal = {Physics of Plasmas},
    volume = {30},
    number = {6},
    pages = {062703},
    year = {2023},
    month = {06},
    issn = {1070-664X},
    doi = {10.1063/5.0143738},
    url = {https://doi.org/10.1063/5.0143738}
    }

@article{Glenzer_PRL_2007,
  title = {Observations of Plasmons in Warm Dense Matter},
  author = {Glenzer, S. H. and Landen, O. L. and Neumayer, P. and Lee, R. W. and Widmann, K. and Pollaine, S. W. and Wallace, R. J. and Gregori, G. and H\"oll, A. and Bornath, T. and Thiele, R. and Schwarz, V. and Kraeft, W.-D. and Redmer, R.},
  journal = {Phys. Rev. Lett.},
  volume = {98},
  issue = {6},
  pages = {065002},
  numpages = {4},
  year = {2007},
  month = {Feb},
  publisher = {American Physical Society},
  doi = {10.1103/PhysRevLett.98.065002},
  url = {https://link.aps.org/doi/10.1103/PhysRevLett.98.065002}
}

@article{HOLL2007120,
title = {{T}homson scattering from near-solid density plasmas using soft X-ray free electron lasers},
journal = {High Energy Density Physics},
volume = {3},
number = {1},
pages = {120-130},
year = {2007},
note = {Radiative Properties of Hot Dense Matter},
issn = {1574-1818},
doi = {https://doi.org/10.1016/j.hedp.2007.02.033},
url = {https://www.sciencedirect.com/science/article/pii/S1574181807000195},
author = {A. Höll and Th. Bornath and L. Cao and T. Döppner and S. Düsterer and E. Förster and C. Fortmann and S.H. Glenzer and G. Gregori and T. Laarmann and K.-H. Meiwes-Broer and A. Przystawik and P. Radcliffe and R. Redmer and H. Reinholz and G. Röpke and R. Thiele and J. Tiggesbäumker and S. Toleikis and N.X. Truong and T. Tschentscher and I. Uschmann and U. Zastrau}
}

@article{Toleikis_2010,
doi = {10.1088/0953-4075/43/19/194017},
url = {https://doi.org/10.1088/0953-4075/43/19/194017},
year = {2010},
month = {sep},
publisher = {},
volume = {43},
number = {19},
pages = {194017},
author = {Toleikis, S and Bornath, T and Döppner, T and Düsterer, S and Fäustlin, R R and Förster, E and Fortmann, C and Glenzer, S H and Göde, S and Gregori, G and Irsig, R and Laarmann, T and Lee, H J and Li, B and Meiwes-Broer, K-H and Mithen, J and Nagler, B and Przystawik, A and Radcliffe, P and Redlin, H and Redmer, R and Reinholz, H and Röpke, G and Tavella, F and Thiele, R and Tiggesbäumker, J and Uschmann, I and Vinko, S M and Whitcher, T and Zastrau, U and Ziaja, B and Tschentscher, T},
title = {Probing near-solid density plasmas using soft x-ray scattering},
journal = {Journal of Physics B: Atomic, Molecular and Optical Physics}
}

@article{Thiele_PRE_2010,
  title = {{T}homson scattering on inhomogeneous targets},
  author = {Thiele, R. and Sperling, P. and Chen, M. and Bornath, Th. and F\"austlin, R. R. and Fortmann, C. and Glenzer, S. H. and Kraeft, W.-D. and Pukhov, A. and Toleikis, S. and Tschentscher, Th. and Redmer, R.},
  journal = {Phys. Rev. E},
  volume = {82},
  issue = {5},
  pages = {056404},
  numpages = {7},
  year = {2010},
  month = {Nov},
  publisher = {American Physical Society},
  doi = {10.1103/PhysRevE.82.056404},
  url = {https://link.aps.org/doi/10.1103/PhysRevE.82.056404}
}

@article{SPERLING2011145,
title = {Two-color {T}homson scattering at {FLASH}},
journal = {High Energy Density Physics},
volume = {7},
number = {3},
pages = {145-149},
year = {2011},
issn = {1574-1818},
doi = {https://doi.org/10.1016/j.hedp.2011.04.001},
url = {https://www.sciencedirect.com/science/article/pii/S1574181811000425},
author = {P. Sperling and R. Thiele and B. Holst and C. Fortmann and S.H. Glenzer and S. Toleikis and Th. Tschentscher and R. Redmer}
}

@article{Ramakrishna_PRB_2021,
  title = {First-principles modeling of plasmons in aluminum under ambient and extreme conditions},
  author = {Ramakrishna, Kushal and Cangi, Attila and Dornheim, Tobias and Baczewski, Andrew and Vorberger, Jan},
  journal = {Phys. Rev. B},
  volume = {103},
  issue = {12},
  pages = {125118},
  numpages = {15},
  year = {2021},
  month = {Mar},
  publisher = {American Physical Society},
  doi = {10.1103/PhysRevB.103.125118},
  url = {https://link.aps.org/doi/10.1103/PhysRevB.103.125118}
}

@article{Chihara_1987,
	doi = {10.1088/0305-4608/17/2/002},
	url = {https://doi.org/10.1088/0305-4608/17/2/002},
	year = 1987,
	month = {feb},
	publisher = {{IOP} Publishing},
	volume = {17},
	number = {2},
	pages = {295--304},
	author = {J Chihara},
	title = {Difference in X-ray scattering between metallic and non-metallic liquids due to conduction electrons},
	journal = {Journal of Physics F: Metal Physics}
}

@article{Baczewski_PRL_2016,
  title = {X-ray {T}homson Scattering in Warm Dense Matter without the {C}hihara Decomposition},
  author = {Baczewski, A. D. and Shulenburger, L. and Desjarlais, M. P. and Hansen, S. B. and Magyar, R. J.},
  journal = {Phys. Rev. Lett.},
  volume = {116},
  issue = {11},
  pages = {115004},
  numpages = {6},
  year = {2016},
  month = {Mar},
  publisher = {American Physical Society},
  doi = {10.1103/PhysRevLett.116.115004},
  url = {https://link.aps.org/doi/10.1103/PhysRevLett.116.115004}
}

@article{Witte_PRL_2017,
  title = {Warm Dense Matter Demonstrating Non-Drude Conductivity from Observations of Nonlinear Plasmon Damping},
  author = {Witte, B. B. L. and Fletcher, L. B. and Galtier, E. and Gamboa, E. and Lee, H. J. and Zastrau, U. and Redmer, R. and Glenzer, S. H. and Sperling, P.},
  journal = {Phys. Rev. Lett.},
  volume = {118},
  issue = {22},
  pages = {225001},
  numpages = {6},
  year = {2017},
  month = {May},
  publisher = {American Physical Society},
  doi = {10.1103/PhysRevLett.118.225001},
  url = {https://link.aps.org/doi/10.1103/PhysRevLett.118.225001}
}

@Article{Pascarelli_NatureReviews_2023,
author={Pascarelli, Sakura
and McMahon, Malcolm
and P{\'e}pin, Charles
and Mathon, Olivier
and Smith, Raymond F.
and Mao, Wendy L.
and Liermann, Hanns-Peter
and Loubeyre, Paul},
title={Materials under extreme conditions using large X-ray facilities},
journal={Nature Reviews Methods Primers},
year={2023},
month={Nov},
day={02},
volume={3},
number={1},
pages={82},
issn={2662-8449},
doi={10.1038/s43586-023-00264-5},
url={https://doi.org/10.1038/s43586-023-00264-5}
}

@article{Martin_POP_2025,
    author = {Martin, W. M. and Nilsen, J. and Fletcher, L. B. and MacDonald, M. J. and Andersen, L. and Arnott, A. and Bellenbaum, H. and Böhme, M. and Boiadjieva, N. and Czapla, N. and Cowan, T. E. and Döppner, T. and Dyer, G. and Ettelbrick, R. N. and Falcone, R. and Faubel, S. and Galtier, E. and Laso Garcia, A. and Gawne, T. and Hancock, L. and Hart, P. and Hartley, N. J. and Herbert, M. L. and Huang, X. and Jain, G. and Kabelitz, K. D. and Khaghani, D. and Kraus, D. and Lee, H. J. and Li, P. and Lu, Y. and Mettry-Yassa, M. and McGehee, P. and Nagler, B. and Rips, J. and Schumacher, S. and Toro, E. R. and Toncian, T. and Xia, X. and Gleason, A. and Glenzer, S. H.},
    title = {Characterizing laser-heated polymer foams with simultaneous x-ray fluorescence spectroscopy and {T}homson scattering at the Matter in Extreme Conditions Endstation at {LCLS}},
    journal = {Physics of Plasmas},
    volume = {32},
    number = {7},
    pages = {072701},
    year = {2025},
    month = {07},
    issn = {1070-664X},
    doi = {10.1063/5.0267033},
    url = {https://doi.org/10.1063/5.0267033}
    }

@article{PhysRevResearch.2.023260,
  title = {Carbon ionization at gigabar pressures: An ab initio perspective on astrophysical high-density plasmas},
  author = {Bethkenhagen, Mandy and Witte, Bastian B. L. and Sch\"orner, Maximilian and R\"opke, Gerd and D\"oppner, Tilo and Kraus, Dominik and Glenzer, Siegfried H. and Sterne, Philip A. and Redmer, Ronald},
  journal = {Phys. Rev. Research},
  volume = {2},
  issue = {2},
  pages = {023260},
  numpages = {7},
  year = {2020},
  month = {Jun},
  publisher = {American Physical Society},
  doi = {10.1103/PhysRevResearch.2.023260},
  url = {https://link.aps.org/doi/10.1103/PhysRevResearch.2.023260}
}

@article{Regan_PRL_2012,
  title = {Inelastic X-Ray Scattering from Shocked Liquid Deuterium},
  author = {Regan, S. P. and Falk, K. and Gregori, G. and Radha, P. B. and Hu, S. X. and Boehly, T. R. and Crowley, B. J. B. and Glenzer, S. H. and Landen, O. L. and Gericke, D. O. and D\"oppner, T. and Meyerhofer, D. D. and Murphy, C. D. and Sangster, T. C. and Vorberger, J.},
  journal = {Phys. Rev. Lett.},
  volume = {109},
  issue = {26},
  pages = {265003},
  numpages = {5},
  year = {2012},
  month = {Dec},
  publisher = {American Physical Society},
  doi = {10.1103/PhysRevLett.109.265003},
  url = {https://link.aps.org/doi/10.1103/PhysRevLett.109.265003}
}

@Article{White_Nature_2025,
author={White, Thomas G.
and Griffin, Travis D.
and Haden, Daniel
and Lee, Hae Ja
and Galtier, Eric
and Cunningham, Eric
and Khaghani, Dimitri
and Descamps, Adrien
and Wollenweber, Lennart
and Armentrout, Ben
and Convery, Carson
and Appel, Karen
and Fletcher, Luke B.
and Goede, Sebastian
and Hastings, J. B.
and Iratcabal, Jeremy
and McBride, Emma E.
and Molina, Jacob
and Monaco, Giulio
and Morrison, Landon
and Stramel, Hunter
and Yunus, Sameen
and Zastrau, Ulf
and Glenzer, Siegfried H.
and Gregori, Gianluca
and Gericke, Dirk O.
and Nagler, Bob},
title={Superheating gold beyond the predicted entropy catastrophe threshold},
journal={Nature},
year={2025},
month={Jul},
day={01},
volume={643},
number={8073},
pages={950-954},
issn={1476-4687},
doi={10.1038/s41586-025-09253-y},
url={https://doi.org/10.1038/s41586-025-09253-y}
}

@misc{bohme2023evidencefreeboundtransitionswarm,
      title={Evidence of free-bound transitions in warm dense matter and their impact on equation-of-state measurements}, 
      author={Maximilian P. Böhme and Luke B. Fletcher and Tilo Döppner and Dominik Kraus and Andrew D. Baczewski and Thomas R. Preston and Michael J. MacDonald and Frank R. Graziani and Zhandos A. Moldabekov and Jan Vorberger and Tobias Dornheim},
      year={2023},
      eprint={2306.17653},
      archivePrefix={arXiv},
      primaryClass={physics.plasm-ph},
      url={https://arxiv.org/abs/2306.17653}, 
}

@ARTICLE{Riley_IEEE_2003,
  author={Riley, D. and Keenan, R. and Topping, S.J. and Khattak, F.Y. and McEvoy, A.-M. and Angulo, J.J. and Lamb, M.J. and Lewis, C.L.S. and Neely, D. and Notley, M.},
  journal={IEEE Transactions on Plasma Science}, 
  title={Potential for {T}homson scatter with an X-ray laser}, 
  year={2003},
  volume={31},
  number={5},
  pages={1016-1022},
  doi={10.1109/TPS.2003.818769}}

@article{LANDEN2001465,
title = {Dense matter characterization by X-ray {T}homson scattering},
journal = {Journal of Quantitative Spectroscopy and Radiative Transfer},
volume = {71},
number = {2},
pages = {465-478},
year = {2001},
note = {Radiative Properties of Hot Dense Matter},
issn = {0022-4073},
doi = {https://doi.org/10.1016/S0022-4073(01)00090-5},
url = {https://www.sciencedirect.com/science/article/pii/S0022407301000905},
author = {O.L. Landen and S.H. Glenzer and M.J. Edwards and R.W. Lee and G.W. Collins and R.C. Cauble and W.W. Hsing and B.A. Hammel}
}

@article{Wollenweber_RSI_2021,
    author = {Wollenweber, L. and Preston, T. R. and Descamps, A. and Cerantola, V. and Comley, A. and Eggert, J. H. and Fletcher, L. B. and Geloni, G. and Gericke, D. O. and Glenzer, S. H. and Göde, S. and Hastings, J. and Humphries, O. S. and Jenei, A. and Karnbach, O. and Konopkova, Z. and Loetzsch, R. and Marx-Glowna, B. and McBride, E. E. and McGonegle, D. and Monaco, G. and Ofori-Okai, B. K. and Palmer, C. A. J. and Plückthun, C. and Redmer, R. and Strohm, C. and Thorpe, I. and Tschentscher, T. and Uschmann, I. and Wark, J. S. and White, T. G. and Appel, K. and Gregori, G. and Zastrau, U.},
    title = {High-resolution inelastic x-ray scattering at the high energy density scientific instrument at the {E}uropean {X}-Ray {F}ree-{E}lectron {L}aser},
    journal = {Review of Scientific Instruments},
    volume = {92},
    number = {1},
    pages = {013101},
    year = {2021},
    month = {01},
    issn = {0034-6748},
    doi = {10.1063/5.0022886},
    url = {https://doi.org/10.1063/5.0022886}
    }

@article{KRITCHER2011271,
title = {Development of X-ray {T}homson scattering for implosion target characterization},
journal = {High Energy Density Physics},
volume = {7},
number = {4},
pages = {271-276},
year = {2011},
issn = {1574-1818},
doi = {https://doi.org/10.1016/j.hedp.2011.05.013},
url = {https://www.sciencedirect.com/science/article/pii/S1574181811000644},
author = {A.L. Kritcher and T. Döppner and C. Fortmann and O.L. Landen and R. Wallace and S.H. Glenzer}
}

@article{Doeppner_RSI_2016,
    author = {Döppner, T. and Kraus, D. and Neumayer, P. and Bachmann, B. and Emig, J. and Falcone, R. W. and Fletcher, L. B. and Hardy, M. and Kalantar, D. H. and Kritcher, A. L. and Landen, O. L. and Ma, T. and Saunders, A. M. and Wood, R. D.},
    title = {Improving a high-efficiency, gated spectrometer for x-ray {T}homson scattering experiments at the {N}ational {I}gnition {F}acility},
    journal = {Review of Scientific Instruments},
    volume = {87},
    number = {11},
    pages = {11E515},
    year = {2016},
    month = {08},
    issn = {0034-6748},
    doi = {10.1063/1.4959874},
    url = {https://doi.org/10.1063/1.4959874}
    }

@article{Schoerner_PRE_2023,
  title = {X-ray {T}homson scattering spectra from density functional theory molecular dynamics simulations based on a modified {C}hihara formula},
  author = {Sch\"orner, Maximilian and Bethkenhagen, Mandy and D\"oppner, Tilo and Kraus, Dominik and Fletcher, Luke B. and Glenzer, Siegfried H. and Redmer, Ronald},
  journal = {Phys. Rev. E},
  volume = {107},
  issue = {6},
  pages = {065207},
  numpages = {14},
  year = {2023},
  month = {Jun},
  publisher = {American Physical Society},
  doi = {10.1103/PhysRevE.107.065207},
  url = {https://link.aps.org/doi/10.1103/PhysRevE.107.065207}
}

@article{Moldabekov_PRR_2024,
  title = {Excitation signatures of isochorically heated electrons in solids at finite wave number explored from first principles},
  author = {Moldabekov, Zhandos A. and Gawne, Thomas D. and Schwalbe, Sebastian and Preston, Thomas R. and Vorberger, Jan and Dornheim, Tobias},
  journal = {Phys. Rev. Res.},
  volume = {6},
  issue = {2},
  pages = {023219},
  numpages = {11},
  year = {2024},
  month = {May},
  publisher = {American Physical Society},
  doi = {10.1103/PhysRevResearch.6.023219},
  url = {https://link.aps.org/doi/10.1103/PhysRevResearch.6.023219}
}

@article{Moldabekov_MRE_2025,
    author = {Moldabekov, Zhandos A. and Schwalbe, Sebastian and Gawne, Thomas and Preston, Thomas R. and Vorberger, Jan and Dornheim, Tobias},
    title = {Applying the Liouville–Lanczos method of time-dependent density-functional theory to warm dense matter},
    journal = {Matter and Radiation at Extremes},
    volume = {10},
    number = {4},
    pages = {047601},
    year = {2025},
    month = {05},
    issn = {2468-2047},
    doi = {10.1063/5.0263947},
    url = {https://doi.org/10.1063/5.0263947}
    }

@misc{kozlowski2023generalizedgradientapproximationthermal,
      title={Generalized Gradient Approximation Made Thermal}, 
      author={John Kozlowski and Dennis Perchak and Kieron Burke},
      year={2023},
      eprint={2308.03319},
      archivePrefix={arXiv},
      primaryClass={physics.chem-ph},
      url={https://arxiv.org/abs/2308.03319}, 
}

@article{Dornheim_JCP_2023,
    author = {Dornheim, Tobias and Tolias, Panagiotis and Groth, Simon and Moldabekov, Zhandos A. and Vorberger, Jan and Hirshberg, Barak},
    title = {Fermionic physics from ab initio path integral {M}onte {C}arlo simulations of fictitious identical particles},
    journal = {The Journal of Chemical Physics},
    volume = {159},
    number = {16},
    pages = {164113},
    year = {2023},
    month = {10},
    issn = {0021-9606},
    doi = {10.1063/5.0171930},
    url = {https://doi.org/10.1063/5.0171930}
    }

@article{Xiong_JCP_2022,
    author = {Xiong, Yunuo and Xiong, Hongwei},
    title = {On the thermodynamic properties of fictitious identical particles and the application to fermion sign problem},
    journal = {The Journal of Chemical Physics},
    volume = {157},
    number = {9},
    pages = {094112},
    year = {2022},
    month = {09},
    issn = {0021-9606},
    doi = {10.1063/5.0106067},
    url = {https://doi.org/10.1063/5.0106067}
    }

@article{Hilleke_PRM_2025,
  title = {Fully thermal meta-{GGA} exchange correlation free-energy density functional},
  author = {Hilleke, Katerina P. and Karasiev, Valentin V. and Trickey, S. B. and Goshadze, R. M. N. and Hu, S. X.},
  journal = {Phys. Rev. Mater.},
  volume = {9},
  issue = {5},
  pages = {L050801},
  numpages = {8},
  year = {2025},
  month = {May},
  publisher = {American Physical Society},
  doi = {10.1103/PhysRevMaterials.9.L050801},
  url = {https://link.aps.org/doi/10.1103/PhysRevMaterials.9.L050801}
}

@article{moldabekov2025enhancingefficiencytimedependentdensity,
author={Moldabekov, Zhandos A.
and Schwalbe, Sebastian
and Acosta, Uwe Hernandez
and Gawne, Thomas
and Vorberger, Jan
and Pavanello, Michele
and Dornheim, Tobias},
title={Enhancing the efficiency of time-dependent density functional theory calculations of dynamic response properties},
journal={npj Computational Materials},
year={2026},
month={Apr},
day={25},
issn={2057-3960},
doi={10.1038/s41524-026-02088-9},
url={https://doi.org/10.1038/s41524-026-02088-9}
}

@Article{Brygoo2021,
author={Brygoo, S.
and Loubeyre, P.
and Millot, M.
and Rygg, J. R.
and Celliers, P. M.
and Eggert, J. H.
and Jeanloz, R.
and Collins, G. W.},
title={Evidence of hydrogen-helium immiscibility at {J}upiter-interior conditions},
journal={Nature},
year={2021},
month={May},
day={01},
volume={593},
number={7860},
pages={517-521},
issn={1476-4687},
doi={10.1038/s41586-021-03516-0},
url={https://doi.org/10.1038/s41586-021-03516-0}
}

@Article{Moldabekov_ACS_OMEGA_2024,
author={Moldabekov, Zhandos
and Gawne, Thomas D.
and Schwalbe, Sebastian
and Preston, Thomas R.
and Vorberger, Jan
and Dornheim, Tobias},
title={Ultrafast Heating-Induced Suppression of d-Band Dominance in the Electronic Excitation Spectrum of Cuprum},
journal={ACS Omega},
year={2024},
month={Jun},
day={11},
publisher={American Chemical Society},
volume={9},
number={23},
pages={25239-25250},
doi={10.1021/acsomega.4c02920},
url={https://doi.org/10.1021/acsomega.4c02920}
}

@article{Celliers_Science_2018,
author = {Peter M. Celliers  and Marius Millot  and Stephanie Brygoo  and R. Stewart McWilliams  and Dayne E. Fratanduono  and J. Ryan Rygg  and Alexander F. Goncharov  and Paul Loubeyre  and Jon H. Eggert  and J. Luc Peterson  and Nathan B. Meezan  and Sebastien Le Pape  and Gilbert W. Collins  and Raymond Jeanloz  and Russell J. Hemley },
title = {Insulator-metal transition in dense fluid deuterium},
journal = {Science},
volume = {361},
number = {6403},
pages = {677-682},
year = {2018},
doi = {10.1126/science.aat0970},
URL = {https://www.science.org/doi/abs/10.1126/science.aat0970}}

@article{Saunders_PRE_2018,
  title = {Characterizing plasma conditions in radiatively heated solid-density samples with x-ray {T}homson scattering},
  author = {Saunders, A. M. and Lahmann, B. and Sutcliffe, G. and Frenje, J. A. and Falcone, R. W. and D\"oppner, T.},
  journal = {Phys. Rev. E},
  volume = {98},
  issue = {6},
  pages = {063206},
  numpages = {9},
  year = {2018},
  month = {Dec},
  publisher = {American Physical Society},
  doi = {10.1103/PhysRevE.98.063206},
  url = {https://link.aps.org/doi/10.1103/PhysRevE.98.063206}
}

@Article{Smid_SciRep_2026,
author={{\v{S}}m{\'i}d, M.
and Humphries, O. S.
and Baehtz, C.
and Bouffetier, V.
and Brambrink, E.
and Burian, T.
and Cerantola, V.
and Cho, M. S.
and Cowan, T. E.
and Gaus, L.
and Gu, M. F.
and H{\'a}jkov{\'a}, V.
and Juha, L.
and Kaa, J.
and Konopkova, Z.
and Kozlov{\'a}, M.
and Le, H. P.
and Makita, M.
and Pan, X.
and Preston, T. R.
and Schropp, A.
and Schwinkendorf, J.-P.
and Scott, H. A.
and {\v{S}}tefan{\'i}kov{\'a}, R.
and Vorberger, J.
and Wang, W.
and Zastrau, U.
and Falk, K.},
title={Plasma screening in mid-charged ions observed by {K}-shell line emission},
journal={Scientific Reports},
year={2026},
month={Feb},
day={10},
volume={16},
number={1},
pages={5873},
issn={2045-2322},
doi={10.1038/s41598-026-39041-1},
url={https://doi.org/10.1038/s41598-026-39041-1}
}

@article{Lu_Spec_2017,
title = {Characterization of x-ray imaging crystal spectrometer for high-resolution spatially-resolved x-ray {T}homson scattering measurements in shock-compressed experiments},
journal = {Journal of Quantitative Spectroscopy and Radiative Transfer},
volume = {187},
pages = {247-254},
year = {2017},
issn = {0022-4073},
doi = {https://doi.org/10.1016/j.jqsrt.2016.10.001},
url = {https://www.sciencedirect.com/science/article/pii/S0022407316302898},
author = {J. Lu and K.W. Hill and M. Bitter and N.A. Pablant and L.F. Delgado-Aparicio and P.C. Efthimion and H.J. Lee and U. Zastrau}
}

@article{Davis_Instrumentation_2012,
doi = {10.1088/1748-0221/7/02/P02004},
url = {https://doi.org/10.1088/1748-0221/7/02/P02004},
year = {2012},
month = {feb},
publisher = {},
volume = {7},
number = {02},
pages = {P02004},
author = {P Davis and T Döppner and S H Glenzer and R W Falcone and W Unites},
title = {An apparatus for the characterization of warm, dense deuterium with inelastic x-ray scattering},
journal = {Journal of Instrumentation}
}

@article{Lahmann_PPCF_2023,
doi = {10.1088/1361-6587/ace4f2},
url = {https://doi.org/10.1088/1361-6587/ace4f2},
year = {2023},
month = {jul},
publisher = {IOP Publishing},
volume = {65},
number = {9},
pages = {095002},
author = {Lahmann, B and Saunders, A M and Döppner, T and Frenje, J A and Glenzer, S H and Gatu-Johnson, M and Sutcliffe, G and Zylstra, A B and Petrasso, R D},
title = {Measuring stopping power in warm dense matter plasmas at {OMEGA}},
journal = {Plasma Physics and Controlled Fusion}
}

@article{Johnson_PRE_2016,
  title = {Average-atom treatment of relaxation time in x-ray {T}homson scattering from warm dense matter},
  author = {Johnson, W. R. and Nilsen, J.},
  journal = {Phys. Rev. E},
  volume = {93},
  issue = {3},
  pages = {033205},
  numpages = {7},
  year = {2016},
  month = {Mar},
  publisher = {American Physical Society},
  doi = {10.1103/PhysRevE.93.033205},
  url = {https://link.aps.org/doi/10.1103/PhysRevE.93.033205}
}

@article{Zastrau_JSynchRad_2021,
author = "Zastrau, Ulf and Appel, Karen and Baehtz, Carsten and Baehr, Oliver and Batchelor, Lewis and Bergh{\"{a}}user, Andreas and Banjafar, Mohammadreza and Brambrink, Erik and Cerantola, Valerio and Cowan, Thomas E. and Damker, Horst and Dietrich, Steffen and Di Dio Cafiso, Samuele and Dreyer, J{\"{o}}rn and Engel, Hans-Olaf and Feldmann, Thomas and Findeisen, Stefan and Foese, Manon and Fulla-Marsa, Daniel and G{\"{o}}de, Sebastian and Hassan, Mohammed and Hauser, Jens and Herrmannsd{\"{o}}rfer, Thomas and H{\"{o}}ppner, Hauke and Kaa, Johannes and Kaever, Peter and Kn{\"{o}}fel, Klaus and Kon{\^{o}}pkov{\'{a}}, Zuzana and Laso Garc{\'\i}a, Alejandro and Liermann, Hanns-Peter and Mainberger, Jona and Makita, Mikako and Martens, Eike-Christian and McBride, Emma E. and M{\"{o}}ller, Dominik and Nakatsutsumi, Motoaki and Pelka, Alexander and Plueckthun, Christian and Prescher, Clemens and Preston, Thomas R. and R{\"{o}}per, Michael and Schmidt, Andreas and Seidel, Wolfgang and Schwinkendorf, Jan-Patrick and Schoelmerich, Markus O. and Schramm, Ulrich and Schropp, Andreas and Strohm, Cornelius and Sukharnikov, Konstantin and Talkovski, Peter and Thorpe, Ian and Toncian, Monika and Toncian, Toma and Wollenweber, Lennart and Yamamoto, Shingo and Tschentscher, Thomas",
title = {The High Energy Density Scientific Instrument at the {E}uropean {XFEL}},
journal = "Journal of Synchrotron Radiation",
year = "2021",
volume = "28",
number = "5",
pages = "1393--1416",
month = "Sep",
doi = {10.1107/S1600577521007335},
url = {https://doi.org/10.1107/S1600577521007335},
}

@article{Kraus_HEDP_2012,
title = {X-ray {T}homson scattering on shocked graphite},
journal = {High Energy Density Physics},
volume = {8},
number = {1},
pages = {46-49},
year = {2012},
issn = {1574-1818},
doi = {https://doi.org/10.1016/j.hedp.2011.11.011},
url = {https://www.sciencedirect.com/science/article/pii/S1574181811001078},
author = {D. Kraus and A. Otten and A. Frank and V. Bagnoud and A. Blažević and D.O. Gericke and G. Gregori and A. Ortner and G. Schaumann and D. Schumacher and J. Vorberger and F. Wagner and K. Wünsch and M. Roth}
}

@article{Fletcher_POP_2013,
    author = {Fletcher, L. B. and Kritcher, A. and Pak, A. and Ma, T. and Döppner, T. and Fortmann, C. and Divol, L. and Landen, O. L. and Vorberger, J. and Chapman, D. A. and Gericke, D. O. and Falcone, R. W. and Glenzer, S. H.},
    title = {X-ray {T}homson scattering measurements of temperature and density from multi-shocked {CH} capsules)},
    journal = {Physics of Plasmas},
    volume = {20},
    number = {5},
    pages = {056316},
    year = {2013},
    month = {05},
    issn = {1070-664X},
    doi = {10.1063/1.4807032},
    url = {https://doi.org/10.1063/1.4807032}
    }

@article{Mattern_PRB_2012,
  title = {Real-space Green's function calculations of Compton profiles},
  author = {Mattern, Brian A. and Seidler, Gerald T. and Kas, Joshua J. and Pacold, Joseph I. and Rehr, John J.},
  journal = {Phys. Rev. B},
  volume = {85},
  issue = {11},
  pages = {115135},
  numpages = {10},
  year = {2012},
  month = {Mar},
  publisher = {American Physical Society},
  doi = {10.1103/PhysRevB.85.115135},
  url = {https://link.aps.org/doi/10.1103/PhysRevB.85.115135}
}

@article{Fletcher_PRL_2014,
  title = {Observations of Continuum Depression in Warm Dense Matter with X-Ray {T}homson Scattering},
  author = {Fletcher, L. B. and Kritcher, A. L. and Pak, A. and Ma, T. and D\"oppner, T. and Fortmann, C. and Divol, L. and Jones, O. S. and Landen, O. L. and Scott, H. A. and Vorberger, J. and Chapman, D. A. and Gericke, D. O. and Mattern, B. A. and Seidler, G. T. and Gregori, G. and Falcone, R. W. and Glenzer, S. H.},
  journal = {Phys. Rev. Lett.},
  volume = {112},
  issue = {14},
  pages = {145004},
  numpages = {5},
  year = {2014},
  month = {Apr},
  publisher = {American Physical Society},
  doi = {10.1103/PhysRevLett.112.145004},
  url = {https://link.aps.org/doi/10.1103/PhysRevLett.112.145004}
}

@book{Riley_book,
author = {Riley, David},
title = {Warm Dense Matter},
publisher = {IOP Publishing},
year = {2021},
series = {2053-2563},
isbn = {978-0-7503-2348-2},
url = {https://doi.org/10.1088/978-0-7503-2348-2},
doi = {10.1088/978-0-7503-2348-2}
}

@article{Hesselbach_MRE_2024,
    author = {Hesselbach, P. and Lütgert, J. and Bagnoud, V. and Belikov, R. and Humphries, O. and Lindqvist, B. and Schaumann, G. and Sokolov, A. and Tauschwitz, A. and Varentsov, D. and Weyrich, K. and Winkler, B. and Yu, X. and Zielbauer, B. and Kraus, D. and Riley, D. and Major, Zs. and Neumayer, P.},
    title = {Platform for laser-driven X-ray diagnostics of heavy-ion heated extreme states of matter},
    journal = {Matter and Radiation at Extremes},
    volume = {10},
    number = {1},
    pages = {017803},
    year = {2024},
    month = {12},
    issn = {2468-2047},
    doi = {10.1063/5.0233548},
    url = {https://doi.org/10.1063/5.0233548}
    }

@article{Falk_JPhysConf_2010,
doi = {10.1088/1742-6596/244/4/042014},
url = {https://doi.org/10.1088/1742-6596/244/4/042014},
year = {2010},
month = {aug},
publisher = {},
volume = {244},
number = {4},
pages = {042014},
author = {K Falk and A P Jephcoat and B J B Crowley and R R Fäustlin and C Fortmann and F Y Khattak and A K Kleppe and D Riley and S Toleikis and J Wark and H Wilhelm and G Gregori},
title = {Measurement of the dynamic response of compressed hydrogen by inelastic X-ray scattering},
journal = {Journal of Physics: Conference Series}
}

@article{Falk_PPCF_2020,
doi = {10.1088/1361-6587/ab8bb3},
url = {https://doi.org/10.1088/1361-6587/ab8bb3},
year = {2020},
month = {may},
publisher = {IOP Publishing},
volume = {62},
number = {7},
pages = {074001},
author = {Falk, K and Fontes, C J and Fryer, C L and Greeff, C W and Holec, M and Johns, H M and Montgomery, D S and Schmidt, D W and Šmíd, M},
title = {Experimental observation of elevated heating in dynamically compressed {CH} foam},
journal = {Plasma Physics and Controlled Fusion}
}

@article{Falk_PPCF_2017,
doi = {10.1088/0741-3335/59/1/014050},
url = {https://doi.org/10.1088/0741-3335/59/1/014050},
year = {2016},
month = {nov},
publisher = {IOP Publishing},
volume = {59},
number = {1},
pages = {014050},
author = {Falk, K and Fryer, C L and Gamboa, E J and Greeff, C W and Johns, H M and Schmidt, D W and Šmíd, M and Benage, J F and Montgomery, D S},
title = {X-ray {T}homson scattering measurement of temperature in warm dense carbon},
journal = {Plasma Physics and Controlled Fusion}
}

@article{LePape_POP_2010,
    author = {Le Pape, Sebastien and Neumayer, Paul and Fortmann, Carsten and Döppner, Tilo and Davis, Paul and Kritcher, Andrea and Landen, Otto and Glenzer, Siegfried},
    title = {X-ray radiography and scattering diagnosis of dense shock-compressed matter},
    journal = {Physics of Plasmas},
    volume = {17},
    number = {5},
    pages = {056309},
    year = {2010},
    month = {04},
    issn = {1070-664X},
    doi = {10.1063/1.3377785},
    url = {https://doi.org/10.1063/1.3377785}
    }

@article{Gamboa_HEDP_2014,
title = {Spatially-resolved X-ray scattering measurements of a planar blast wave},
journal = {High Energy Density Physics},
volume = {11},
pages = {75-79},
year = {2014},
issn = {1574-1818},
doi = {https://doi.org/10.1016/j.hedp.2014.03.002},
url = {https://www.sciencedirect.com/science/article/pii/S157418181400024X},
author = {E.J. Gamboa and P.A. Keiter and R.P. Drake and K. Falk and D.S. Montgomery and J.F. Benage}
}

@article{Clerouin_PRE_2015,
  title = {Evidence for out-of-equilibrium states in warm dense matter probed by x-ray {T}homson scattering},
  author = {Cl\'erouin, Jean and Robert, Gr\'egory and Arnault, Philippe and Ticknor, Christopher and Kress, Joel D. and Collins, Lee A.},
  journal = {Phys. Rev. E},
  volume = {91},
  issue = {1},
  pages = {011101},
  numpages = {5},
  year = {2015},
  month = {Jan},
  publisher = {American Physical Society},
  doi = {10.1103/PhysRevE.91.011101},
  url = {https://link.aps.org/doi/10.1103/PhysRevE.91.011101}
}

@article{Plagemann_NJP_2012,
doi = {10.1088/1367-2630/14/5/055020},
url = {https://doi.org/10.1088/1367-2630/14/5/055020},
year = {2012},
month = {may},
publisher = {IOP Publishing},
volume = {14},
number = {5},
pages = {055020},
author = {Plagemann, K-U and Sperling, P and Thiele, R and Desjarlais, M P and Fortmann, C and Döppner, T and Lee, H J and Glenzer, S H and Redmer, R},
title = {Dynamic structure factor in warm dense beryllium},
journal = {New Journal of Physics}
}

@article{LePape_NJP_2013,
doi = {10.1088/1367-2630/15/8/085011},
url = {https://doi.org/10.1088/1367-2630/15/8/085011},
year = {2013},
month = {aug},
publisher = {IOP Publishing},
volume = {15},
number = {8},
pages = {085011},
author = {Pape, Sebastien Le and Correa, Alfredo A and Fortmann, Carsten and Neumayer, Paul and Döppner, Tilo and Davis, Paul and Ma, Tammy and Divol, Laurent and Plagemann, Kai-Uwe and Schwegler, E and Redmer, Ronald and Glenzer, Siegfried},
title = {Structure measurements of compressed liquid boron at megabar pressures},
journal = {New Journal of Physics}
}

@article{Gamboa_POP_2014,
    author = {Gamboa, E. J. and Drake, R. P. and Falk, K. and Keiter, P. A. and Montgomery, D. S. and Benage, J. F. and Trantham, M. R.},
    title = {Simultaneous measurements of several state variables in shocked carbon by imaging x-ray scattering},
    journal = {Physics of Plasmas},
    volume = {21},
    number = {4},
    pages = {042701},
    year = {2014},
    month = {04},
    issn = {1070-664X},
    doi = {10.1063/1.4869241},
    url = {https://doi.org/10.1063/1.4869241}
    }

@article{Sawada_POP_2007,
    author = {Sawada, H. and Regan, S. P. and Meyerhofer, D. D. and Igumenshchev, I. V. and Goncharov, V. N. and Boehly, T. R. and Epstein, R. and Sangster, T. C. and Smalyuk, V. A. and Yaakobi, B. and Gregori, G. and Glenzer, S. H. and Landen, O. L.},
    title = {Diagnosing direct-drive, shock-heated, and compressed plastic planar foils with noncollective spectrally resolved x-ray scattering},
    journal = {Physics of Plasmas},
    volume = {14},
    number = {12},
    pages = {122703},
    year = {2007},
    month = {12},
    issn = {1070-664X},
    doi = {10.1063/1.2819675},
    url = {https://doi.org/10.1063/1.2819675}}

@article{Gamboa_POP_2015,
    author = {Gamboa, E. J. and Fletcher, L. B. and Lee, H. J. and Zastrau, U. and Galtier, E. and MacDonald, M. J. and Gauthier, M. and Vorberger, J. and Gericke, D. O. and Granados, E. and Hastings, J. B. and Glenzer, S. H.},
    title = {Single-shot measurements of plasmons in compressed diamond with an x-ray laser},
    journal = {Physics of Plasmas},
    volume = {22},
    number = {5},
    pages = {056319},
    year = {2015},
    month = {05},
    issn = {1070-664X},
    doi = {10.1063/1.4921407},
    url = {https://doi.org/10.1063/1.4921407}
    }

@article{Lv_POP_2019,
    author = {Lv, Min and Hu, Zhimin and Hou, Yong and Wei, Minxi and Mo, Chongjie and Zheng, Wei and Lv, Meng and Yang, Guohong and Zhao, Yang and Zhang, Zhiyu and Qing, Bo and Xiong, Gang and Zhan, Xiayu and Hou, Lifei and Zhang, Wenhai and Kang, Wei and Zhang, Ping and Yuan, Jianmin and Zhang, Jiyan and Yang, Jiamin},
    title = {Measurement of ionic structure in isochorically heated graphite from X-ray {T}homson scattering},
    journal = {Physics of Plasmas},
    volume = {26},
    number = {2},
    pages = {022702},
    year = {2019},
    month = {02},
    issn = {1070-664X},
    doi = {10.1063/1.5054088},
    url = {https://doi.org/10.1063/1.5054088}
    }

@article{Kraus_POP_2015,
    author = {Kraus, D. and Vorberger, J. and Helfrich, J. and Gericke, D. O. and Bachmann, B. and Bagnoud, V. and Barbrel, B. and Blažević, A. and Carroll, D. C. and Cayzac, W. and Döppner, T. and Fletcher, L. B. and Frank, A. and Frydrych, S. and Gamboa, E. J. and Gauthier, M. and Göde, S. and Granados, E. and Gregori, G. and Hartley, N. J. and Kettle, B. and Lee, H. J. and Nagler, B. and Neumayer, P. and Notley, M. M. and Ortner, A. and Otten, A. and Ravasio, A. and Riley, D. and Roth, F. and Schaumann, G. and Schumacher, D. and Schumaker, W. and Siegenthaler, K. and Spindloe, C. and Wagner, F. and Wünsch, K. and Glenzer, S. H. and Roth, M. and Falcone, R. W.},
    title = {The complex ion structure of warm dense carbon measured by spectrally resolved x-ray scattering},
    journal = {Physics of Plasmas},
    volume = {22},
    number = {5},
    pages = {056307},
    year = {2015},
    month = {05},
    issn = {1070-664X},
    doi = {10.1063/1.4920943},
    url = {https://doi.org/10.1063/1.4920943}
    }

@article{Gregori_AIP_2002,
    author = {Gregori, G. and Glenzer, S. H. and Lee, R. W. and Hicks, D. G. and Pasley, J. and Collins, G. W. and Celliers, P. and Bastea, M. and Eggert, J. and Pollaine, S. M. and Landen, O. L.},
    title = {Calculations and measurements of x‐ray {T}homson scattering spectra in warm dense matter},
    journal = {AIP Conference Proceedings},
    volume = {645},
    number = {1},
    pages = {359-368},
    year = {2002},
    month = {12},
    issn = {0094-243X},
    doi = {10.1063/1.1525476},
    url = {https://doi.org/10.1063/1.1525476}
    }

@article{White_ElectronicStructure_2025,
doi = {10.1088/2516-1075/adad24},
url = {https://doi.org/10.1088/2516-1075/adad24},
year = {2025},
month = {feb},
publisher = {IOP Publishing},
volume = {7},
number = {1},
pages = {014001},
author = {White, Alexander J},
title = {Dynamical structure factors of warm dense matter from time-dependent orbital-free and mixed-stochastic-deterministic density functional theory},
journal = {Electronic Structure}
}

@article{Fortmann_PRL_2012,
  title = {Measurement of the Adiabatic Index in Be Compressed by Counterpropagating Shocks},
  author = {Fortmann, C. and Lee, H. J. and D\"oppner, T. and Falcone, R. W. and Kritcher, A. L. and Landen, O. L. and Glenzer, S. H.},
  journal = {Phys. Rev. Lett.},
  volume = {108},
  issue = {17},
  pages = {175006},
  numpages = {5},
  year = {2012},
  month = {Apr},
  publisher = {American Physical Society},
  doi = {10.1103/PhysRevLett.108.175006},
  url = {https://link.aps.org/doi/10.1103/PhysRevLett.108.175006}
}

@article{Neumayer_RSI_2006,
    author = {Neumayer, P. and Gregori, G. and Ravasio, A. and Koenig, M. and Price, D. and Widmann, K. and Bastea, M. and Landen, O. L. and Glenzer, S. H.},
    title = {Solid-density plasma characterization with x-ray scattering on the 200\,{J} {J}anus laser},
    journal = {Review of Scientific Instruments},
    volume = {77},
    number = {10},
    pages = {10F317},
    year = {2006},
    month = {10},
    issn = {0034-6748},
    doi = {10.1063/1.2235648},
    url = {https://doi.org/10.1063/1.2235648}
    }

@article{Kraus_PRL_2013,
  title = {Probing the Complex Ion Structure in Liquid Carbon at 100\,{GPa}},
  author = {Kraus, D. and Vorberger, J. and Gericke, D. O. and Bagnoud, V. and Bla\ifmmode \check{z}\else \v{z}\fi{}evi\ifmmode \acute{c}\else \'{c}\fi{}, A. and Cayzac, W. and Frank, A. and Gregori, G. and Ortner, A. and Otten, A. and Roth, F. and Schaumann, G. and Schumacher, D. and Siegenthaler, K. and Wagner, F. and W\"unsch, K. and Roth, M.},
  journal = {Phys. Rev. Lett.},
  volume = {111},
  issue = {25},
  pages = {255501},
  numpages = {5},
  year = {2013},
  month = {Dec},
  publisher = {American Physical Society},
  doi = {10.1103/PhysRevLett.111.255501},
  url = {https://link.aps.org/doi/10.1103/PhysRevLett.111.255501}
}

@article{Neumayer_PRL_2010,
  title = {Plasmons in Strongly Coupled Shock-Compressed Matter},
  author = {Neumayer, P. and Fortmann, C. and D\"oppner, T. and Davis, P. and Falcone, R. W. and Kritcher, A. L. and Landen, O. L. and Lee, H. J. and Lee, R. W. and Niemann, C. and Le Pape, S. and Glenzer, S. H.},
  journal = {Phys. Rev. Lett.},
  volume = {105},
  issue = {7},
  pages = {075003},
  numpages = {4},
  year = {2010},
  month = {Aug},
  publisher = {American Physical Society},
  doi = {10.1103/PhysRevLett.105.075003},
  url = {https://link.aps.org/doi/10.1103/PhysRevLett.105.075003}
}

@article{Kraus_PRE_2016,
  title = {X-ray scattering measurements on imploding {CH} spheres at the {N}ational {I}gnition {F}acility},
  author = {Kraus, D. and Chapman, D. A. and Kritcher, A. L. and Baggott, R. A. and Bachmann, B. and Collins, G. W. and Glenzer, S. H. and Hawreliak, J. A. and Kalantar, D. H. and Landen, O. L. and Ma, T. and Le Pape, S. and Nilsen, J. and Swift, D. C. and Neumayer, P. and Falcone, R. W. and Gericke, D. O. and D\"oppner, T.},
  journal = {Phys. Rev. E},
  volume = {94},
  issue = {1},
  pages = {011202},
  numpages = {5},
  year = {2016},
  month = {Jul},
  publisher = {American Physical Society},
  doi = {10.1103/PhysRevE.94.011202},
  url = {https://link.aps.org/doi/10.1103/PhysRevE.94.011202}
}

@article{Plagemann_PRE_2015,
  title = {Ab initio calculation of the ion feature in x-ray {T}homson scattering},
  author = {Plagemann, Kai-Uwe and R\"uter, Hannes R. and Bornath, Thomas and Shihab, Mohammed and Desjarlais, Michael P. and Fortmann, Carsten and Glenzer, Siegfried H. and Redmer, Ronald},
  journal = {Phys. Rev. E},
  volume = {92},
  issue = {1},
  pages = {013103},
  numpages = {8},
  year = {2015},
  month = {Jul},
  publisher = {American Physical Society},
  doi = {10.1103/PhysRevE.92.013103},
  url = {https://link.aps.org/doi/10.1103/PhysRevE.92.013103}
}

@article{Gregori_JQuantSpec_2006,
title = {Measurement of carbon ionization balance in high-temperature plasma mixtures by temporally resolved X-ray scattering},
journal = {Journal of Quantitative Spectroscopy and Radiative Transfer},
volume = {99},
number = {1},
pages = {225-237},
year = {2006},
issn = {0022-4073},
doi = {https://doi.org/10.1016/j.jqsrt.2005.05.017},
url = {https://www.sciencedirect.com/science/article/pii/S0022407305001470},
author = {G. Gregori and S.H. Glenzer and H.-K. Chung and D.H. Froula and R.W. Lee and N.B. Meezan and J.D. Moody and C. Niemann and O.L. Landen and B. Holst and R. Redmer and S.P. Regan and H. Sawada}
}

@Article{Valenzuela_SciRep_2018,
author={Valenzuela, J. C.
and Krauland, C.
and Mariscal, D.
and Krasheninnikov, I.
and Niemann, C.
and Ma, T.
and Mabey, P.
and Gregori, G.
and Wiewior, P.
and Covington, A. M.
and Beg, F. N.},
title={Measurement of temperature and density using non-collective X-ray {T}homson scattering in pulsed power produced warm dense plasmas},
journal={Scientific Reports},
year={2018},
month={May},
day={30},
volume={8},
number={1},
pages={8432},
issn={2045-2322},
doi={10.1038/s41598-018-26608-w},
url={https://doi.org/10.1038/s41598-018-26608-w}
}

@article{Dopp,
doi = {10.1088/1742-6596/244/3/032044},
url = {https://doi.org/10.1088/1742-6596/244/3/032044},
year = {2010},
month = {aug},
publisher = {},
volume = {244},
number = {3},
pages = {032044},
author = {T Döppner and C Fortmann and P F Davis and A L Kritcher and O L Landen and H J Lee and R Redmer and S P Regan and S H Glenzer},
title = {X-ray {T}homson scattering for measuring dense beryllium plasma collisionality},
journal = {Journal of Physics: Conference Series}
}

@article{Gorman_POP_2024,
    author = {Gorman, M. G. and McGonegle, D. and Smith, R. F. and Singh, S. and Jenkins, T. and McWilliams, R. S. and Albertazzi, B. and Ali, S. J. and Antonelli, L. and Armstrong, M. R. and Baehtz, C. and Ball, O. B. and Banerjee, S. and Belonoshko, A. B. and Benuzzi-Mounaix, A. and Bolme, C. A. and Bouffetier, V. and Briggs, R. and Buakor, K. and Butcher, T. and Di Dio Cafiso, S. and Cerantola, V. and Chantel, J. and Di Cicco, A. and Clarke, S. and Coleman, A. L. and Collier, J. and Collins, G. W. and Comley, A. J. and Coppari, F. and Cowan, T. E. and Cristoforetti, G. and Cynn, H. and Descamps, A. and Dorchies, F. and Duff, M. J. and Dwivedi, A. and Edwards, C. and Eggert, J. H. and Errandonea, D. and Fiquet, G. and Galtier, E. and Laso Garcia, A. and Ginestet, H. and Gizzi, L. and Gleason, A. and Goede, S. and Gonzalez, J. M. and Harmand, M. and Hartley, N. J. and Heighway, P. G. and Hernandez-Gomez, C. and Higginbotham, A. and Höppner, H. and Husband, R. J. and Hutchinson, T. M. and Hwang, H. and Lazicki, A. E. and Keen, D. A. and Kim, J. and Koester, P. and Konopkova, Z. and Kraus, D. and Krygier, A. and Labate, L. and Lee, Y. and Liermann, H.-P. and Mason, P. and Masruri, M. and Massani, B. and McBride, E. E. and McGuire, C. and McHardy, J. D. and Merkel, S. and Morard, G. and Nagler, B. and Nakatsutsumi, M. and Nguyen-Cong, K. and Norton, A.-M. and Oleynik, I. I. and Otzen, C. and Ozaki, N. and Pandolfi, S. and Peake, D. J. and Pelka, A. and Pereira, K. A. and Phillips, J. P. and Prescher, C. and Preston, T. R. and Randolph, L. and Ranjan, D. and Ravasio, A. and Redmer, R. and Rips, J. and Santamaria-Perez, D. and Savage, D. J. and Schoelmerich, M. and Schwinkendorf, J.-P. and Smith, J. and Sollier, A. and Spear, J. and Spindloe, C. and Stevenson, M. and Strohm, C. and Suer, T.-A. and Tang, M. and Toncian, M. and Toncian, T. and Tracy, S. J. and Trapananti, A. and Tschentscher, T. and Tyldesley, M. and Vennari, C. E. and Vinci, T. and Vogel, S. C. and Volz, T. J. and Vorberger, J. and Walsh, J. P. S. and Wark, J. S. and Willman, J. T. and Wollenweber, L. and Zastrau, U. and Brambrink, E. and Appel, K. and McMahon, M. I.},
    title = {Shock compression experiments using the DiPOLE 100-X laser on the high energy density instrument at the {E}uropean x-ray free electron laser: Quantitative structural analysis of liquid {Sn}},
    journal = {Journal of Applied Physics},
    volume = {135},
    number = {16},
    pages = {165902},
    year = {2024},
    month = {04},
    issn = {0021-8979},
    doi = {10.1063/5.0201702},
    url = {https://doi.org/10.1063/5.0201702}
    }

@Article{Lazicki2021,
author={Lazicki, A.
and McGonegle, D.
and Rygg, J. R.
and Braun, D. G.
and Swift, D. C.
and Gorman, M. G.
and Smith, R. F.
and Heighway, P. G.
and Higginbotham, A.
and Suggit, M. J.
and Fratanduono, D. E.
and Coppari, F.
and Wehrenberg, C. E.
and Kraus, R. G.
and Erskine, D.
and Bernier, J. V.
and McNaney, J. M.
and Rudd, R. E.
and Collins, G. W.
and Eggert, J. H.
and Wark, J. S.},
title={Metastability of diamond ramp-compressed to 2 terapascals},
journal={Nature},
year={2021},
month={Jan},
day={01},
volume={589},
number={7843},
pages={532-535},
issn={1476-4687},
doi={10.1038/s41586-020-03140-4},
url={https://doi.org/10.1038/s41586-020-03140-4}
}

@article{Harding_RSI_2015,
    author = {Harding, E. C. and Ao, T. and Bailey, J. E. and Loisel, G. and Sinars, D. B. and Geissel, M. and Rochau, G. A. and Smith, I. C.},
    title = {Analysis and implementation of a space resolving spherical crystal spectrometer for x-ray {T}homson scattering experiments},
    journal = {Review of Scientific Instruments},
    volume = {86},
    number = {4},
    pages = {043504},
    year = {2015},
    month = {04},
    issn = {0034-6748},
    doi = {10.1063/1.4918619},
    url = {https://doi.org/10.1063/1.4918619}
    }

@article{Kraus_POP_2018,
    author = {Kraus, D. and Hartley, N. J. and Frydrych, S. and Schuster, A. K. and Rohatsch, K. and Rödel, M. and Cowan, T. E. and Brown, S. and Cunningham, E. and van Driel, T. and Fletcher, L. B. and Galtier, E. and Gamboa, E. J. and Laso Garcia, A. and Gericke, D. O. and Granados, E. and Heimann, P. A. and Lee, H. J. and MacDonald, M. J. and MacKinnon, A. J. and McBride, E. E. and Nam, I. and Neumayer, P. and Pak, A. and Pelka, A. and Prencipe, I. and Ravasio, A. and Redmer, R. and Saunders, A. M. and Schölmerich, M. and Schörner, M. and Sun, P. and Turner, S. J. and Zettl, A. and Falcone, R. W. and Glenzer, S. H. and Döppner, T. and Vorberger, J.},
    title = {High-pressure chemistry of hydrocarbons relevant to planetary interiors and inertial confinement fusion},
    journal = {Physics of Plasmas},
    volume = {25},
    number = {5},
    pages = {056313},
    year = {2018},
    month = {05},
    issn = {1070-664X},
    doi = {10.1063/1.5017908},
    url = {https://doi.org/10.1063/1.5017908}
    }

@article{White_PRR_2024,
  title = {Speed of sound in methane under conditions of planetary interiors},
  author = {White, Thomas G. and Poole, Hannah and McBride, Emma E. and Oliver, Matthew and Descamps, Adrien and Fletcher, Luke B. and Angermeier, W. Alex and Allen, Cameron H. and Appel, Karen and Condamine, Florian P. and Curry, Chandra B. and Dallari, Francesco and Funk, Stefan and Galtier, Eric and Gamboa, Eliseo J. and Gauthier, Maxence and Graham, Peter and Goede, Sebastian and Haden, Daniel and Kim, Jongjin B. and Lee, Hae Ja and Ofori-Okai, Benjamin K. and Richardson, Scott and Rigby, Alex and Schoenwaelder, Christopher and Sun, Peihao and Witte, Bastian L. and Tschentscher, Thomas and Zastrau, Ulf and Nagler, Bob and Hastings, J. B. and Monaco, Giulio and Gericke, Dirk O. and Glenzer, Siegfried H. and Gregori, Gianluca},
  journal = {Phys. Rev. Res.},
  volume = {6},
  issue = {2},
  pages = {L022029},
  numpages = {7},
  year = {2024},
  month = {May},
  publisher = {American Physical Society},
  doi = {10.1103/PhysRevResearch.6.L022029},
  url = {https://link.aps.org/doi/10.1103/PhysRevResearch.6.L022029}
}

@book{quantum_theory,
address = {Cambridge},
author = {G. Giuliani and G. Vignale},
publisher = {Cambridge University Press},
title = {Quantum Theory of the Electron Liquid},
year = {2008},
}

\end{document}